\documentclass[aps, superscriptaddress, showpacs,prx, twocolumn]{revtex4-1}
\usepackage{ORI_Group_style}
\usepackage{blindtext}
\usepackage{mathtools}
\usepackage[dvipsnames]{xcolor}
\usepackage{array}
\usepackage{soul}

\newcolumntype{M}[1]{>{\centering\arraybackslash}m{#1}}

\hypersetup{colorlinks,linkcolor=red,citecolor=blue,urlcolor=blue}

\newcommand{\PoDF}{\mathbf{x}_{\Omega}} 
\newcommand{\QPoDF}{\hat{\mathbf{x}}_{\Omega}} 
\newcommand{\PoDFc}{x_{\Omega}} 
\newcommand{\PoDFs}[1]{\mathbf{x}_{\Omega,#1}} 
\newcommand{\MPo}{m_{\Omega}} 
\newcommand{\FPo}{\Omega} 
\newcommand{\PropPo}{G_{\Omega}} 
\newcommand{\MoPo}{\hat{\mathbf{p}}_{\Omega}} 

\newcommand{\PropNP}{G_{\rm NP}} 
\newcommand{\NoiseNP}{\mathcal{N}_{\rm NP}} 

\newcommand{\PhDF}{\mathbf{x}_{\theta}} 
\newcommand{\QPhDF}{\hat{\mathbf{x}}_{\theta}} 
\newcommand{\PhDFc}{x_{\theta}} 
\newcommand{\PhDFs}[1]{\mathbf{x}_{\theta,#1}} 
\newcommand{\MPh}{m_{\theta}} 
\newcommand{\FPh}{\omega_{\theta}} 
\newcommand{\PropPh}{G_{\theta}} 
\newcommand{\HadPh}{H_{\theta}} 
\newcommand{\SCTP}{\mathbf{J}} 
\newcommand{\SCTPc}{J} 
\newcommand{\CorrPh}{\mathcal{Z}} 
\newcommand{\MoPh}{\hat{\mathbf{p}}_{\theta}} 

\newcommand{\PhBath}{\mathbf{x}_{n}} 
\newcommand{\QPhBath}{\hat{\mathbf{x}}_{n}} 
\newcommand{\PhBaths}[1]{\mathbf{x}_{n,#1}} 
\newcommand{\MBath}{m_{n}} 
\newcommand{\FBath}{\omega_{n}} 
\newcommand{\MoBath}{\hat{\mathbf{p}}_{n}} 

\newcommand{\CPoPh}{g} 

\newcommand{\CAPo}{q} 

\newcommand{\CPhBath}{\kappa_{n}} 

\newcommand{\DissQBM}{G_{\gamma}^{0}}
\newcommand{\DissQBMNS}{G}
\newcommand{\NoiQBM}{H_{\gamma}^{0}}
\newcommand{\NoiQBMNS}{H} 
\newcommand{\SpQBM}{J_{\gamma_{I}}} 

\newcommand{\CDamp}{\gamma_{I}} 

\newcommand{\FfCO}{f} 

\newcommand{\EMCO}{\Lambda_{\rm EM}} 
\newcommand{\EMfCO}{\mathcal{I}} 

\newcommand{\PropEM}{\mathbb{D}_{\mu\nu}} 
\newcommand{\PropScalar}{\mathcal{G}} 
\newcommand{\MKerEMR}{\mathbb{G}_{\rm EM}^{jk}} 
\newcommand{\MKerEMH}{\mathbb{H}_{\rm EM}^{jk}} 
\newcommand{\KerEMR}{G_{\rm EM}} 
\newcommand{\FouKerEMR}{\overline{G}_{\rm EM}} 
\newcommand{\KerEMH}{H_{\rm EM}} 
\newcommand{\FouKerEMH}{\overline{H}_{\rm EM}} 
\newcommand{\DampKRR}{\Gamma^{\rm RR}} 
\newcommand{\DampCRR}{\gamma_{\rm RR}} 

\newcommand{\KerNoise}{\mathcal{N}} 

\newcommand{\HCT}{C_{\theta}} 

\begin{document}

\title{Internal Quantum Dynamics of a Nanoparticle in a Thermal Electromagnetic Field: a Minimal Model}

\author{A. E. Rubio L\'opez}
\email{Adrian.Rubio-Lopez@uibk.ac.at}
\affiliation{Institute for Quantum Optics and Quantum Information of the Austrian Academy of Sciences, A-6020 Innsbruck, Austria.}
\affiliation{Institute for Theoretical Physics, University of Innsbruck, A-6020 Innsbruck, Austria.}

\author{C. Gonzalez-Ballestero}
\email{carlos.gonzalez-ballestero@uibk.ac.at}
\affiliation{Institute for Quantum Optics and Quantum Information of the Austrian Academy of Sciences, A-6020 Innsbruck, Austria.}
\affiliation{Institute for Theoretical Physics, University of Innsbruck, A-6020 Innsbruck, Austria.}

\author{O. Romero-Isart}
\affiliation{Institute for Quantum Optics and Quantum Information of the Austrian Academy of Sciences, A-6020 Innsbruck, Austria.}
\affiliation{Institute for Theoretical Physics, University of Innsbruck, A-6020 Innsbruck, Austria.}

\pacs{44.40.+a, 78.67.Bf, 65.80.-g}

\begin{abstract}

We argue that macroscopic electrodynamics is unsuited to describe the process of radiative thermalization between a levitated nanoparticle in high vacuum and the thermal electromagnetic field. Based on physical arguments, we propose a model to describe such systems beyond the quasi-equilibrium approximation. We use path integral techniques to analytically solve the model and exactly calculate the time evolution of the quantum degrees of freedom of the system.  Free parameters of the microscopic quantum model are determined by matching analytical results to well-known macroscopic response functions. The time evolution of the internal energy of a levitated nanoparticle in a thermal electromagnetic field, as described by our model, qualitatively differs from macroscopic electrodynamics, a prediction that can be experimentally tested.

\end{abstract}

\maketitle

\section{Introduction}

The understanding of the interaction between electromagnetic (EM) fields and matter is a cornerstone of physics. At macroscopic scales, one relies on the theory of macroscopic electrodynamics. This consists of Maxwell's equations and constitutive relations equipped with electric and magnetic susceptibilities that phenomenologically characterize the properties of matter. On the other hand, at microscopic scales, the theory of quantum electrodynamics in the non-relativistic regime, namely quantum optics, can be used to study from first principles the interaction between electromagnetic fields and single atoms. The transition between the two descriptions, however, is a challenging problem since the Schr\"odinger equation for a many-particle system becomes rapidly intractable. 

Recently, a significant amount of research activity has been devoted to levitating nanoobjects in high-vacuum: optical levitation of dielectric nanospheres~\cite{RomeroIsart2010,Chang2010,Barker2010,Li2011,Gieseler2012,Kiesel2013,Asenbaum2013,Jain2016}, magnetic levitation of nanomagnets~\cite{Rusconi2017} and superconducting spheres\cite{RomeroIsart2012}, electrostatic levitation of objects hosting quantum emitters~\cite{Rahman2017}, etc. These experiments offer a new paradigm to study the interaction between electromagnetic fields and matter at the interface between quantum optics and macroscopic electrodynamics.
So far, the research on levitated nanoobjects has been mainly focused on controlling their external degrees of freedom, namely the center-of-mass motion and rotation~\cite{Kuhn2015,Kuhn2017}. However,  one can foresee that it  will soon be possible to probe and control their \emph{internal} degrees of freedom. From a condensed matter point of view, such a possibility might thus provide an ideal platform to test and probe the theory of quantum excitations in solids, as well as their interaction with photons, in an extreme scenario: unclamped matter at the nanometer scale, in high vacuum and out of equilibrium.

\begin{figure}[t] 
	\centering
	\includegraphics[width=\linewidth]{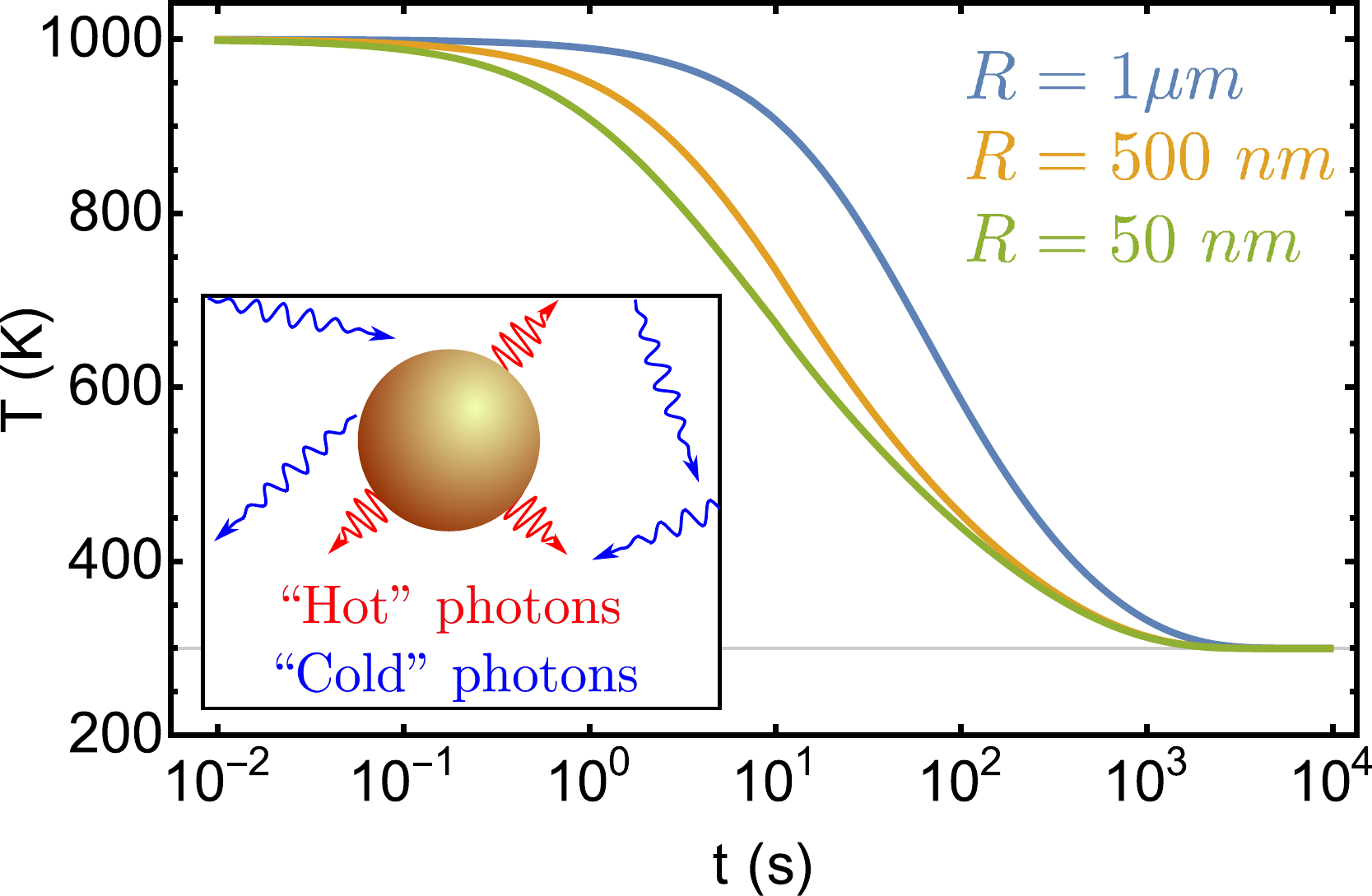}
	\caption{(Color online). Dynamics of radiative thermalization of a gold nanoparticle initially at $T = 1000$ K in a thermal EM environment at $T_{\rm EM} = 300$ K. The temperature dynamics has been calculated using the theory of Fluctuating Electrodynamics (FED), \eqnref{eq:dTFD}, for different nanoparticle radii.}\label{fig1}
\end{figure}

One of the many fundamental questions that may arise in the aforementioned systems is how a levitated nanoobject in high vacuum thermalizes through its interaction with a thermal electromagnetic field, i. e. how its internal energy evolves as a function of time when the system is out of equilibrium \cite{SaavedraACSPhotonics2016}. Within classical macroscopic electrodynamics, this question is answered as follows. A sufficiently small nanoparticle (NP) can be approximated as a point dipole with a frequency-dependent polarizability (assumed scalar for simplicity) given by the Clausius-Mossotti relation \cite{Novotnybook}
\begin{equation}
\alpha (\omega) = 3\epsilon_{0}V\frac{\varepsilon(\omega)-1}{\varepsilon(\omega)+2},
\label{PolNanoSphere}
\end{equation}
where $\epsilon_0$ is the vacuum permeability, $V$ the volume of the nanoparticle, and $\varepsilon(\omega)$ the frequency-dependent dielectric constant. Assuming that the particle has a well defined, spatially homogeneous bulk temperature $T$ at every given time, namely it is in local equilibrium, the fluctuation-dissipation theorem \cite{BiehsBenAbdallah,BenAbdallahPRL2011,ZhuPRB2018,ManjavacasPRB2012,MuletAPL2001} can be used to calculate the power dissipated due to thermal fluctuations of the polarization of the particle,
\be
\mathcal{P}(T)= \hbar\int_{0}^{\infty}\!\!d\omega \chi(\omega)\frac{\omega^{4}}{\pi^2\epsilon_0c^{3}} \spare{n(T,\omega) - n(T_{\rm EM},\omega)}.
\ee
Here $T_\text{EM}$ is the temperature of the electromagnetic 
field, $\chi(\omega)\equiv {\rm Im}[\tilde{\alpha}(\omega)]-\omega^{3}|\tilde{\alpha}(\omega) |^{2}/(6\pi \epsilon_0c^{3})$, $\tilde{\alpha}(\omega) 		= \alpha(\omega)(1-i\alpha(\omega)\omega^3/6\pi c^3)^{-1}$, $ n^{-1}(T,\omega) \equiv\exp[\hbar\omega/(k_{\rm B} T)] -1 $, $c$ is the 		vacuum speed of light, and $k_{\rm B}$ the Boltzmann constant.
To study the thermalization dynamics, one further assumes that the internal equilibration time of the particle is much faster than the rate at which energy is lost due to radiation. Then,
one can conclude that the time evolution of the nanoparticle temperature is governed, over coarse-grained timescales, by the quasi-equilibrium differential equation \cite{BiehsBenAbdallah,ZhuPRB2018}
\begin{equation} \label{eq:dTFD}
\rho V C \frac{dT}{dt}=-\mathcal{P} (T),
\end{equation}
where $\rho $ is the mass density   and $C$ the heat capacity of the nanoparticle. The approach to thermalization described by \eqnref{eq:dTFD} is commonly referred to as Fluctuation Electrodynamics (FED). As an example, we display in \figref{fig1} the thermalization curves of gold nanoparticles with different radii, modeled by a Drude permittivity $\varepsilon(\omega) = 1-\omega_{\rm Pl}^2/(\omega^2+i\omega\gamma_{\rm D})$ with Plasma frequency $\omega_{\rm Pl} = 2\pi\times 2.72\cdot10^{15}$Hz and damping $\gamma_{\rm D} = 2\pi\times6.45\cdot10^{12}$Hz \cite{ChapuisPRB2008}.

In this article, we aim at exploring the radiative thermalization of a levitated nanoparticle in high vacuum, an extreme regime which we think cannot be well described by FED. Our work is divided in four sections: first, we devote Sec.~\ref{SecHeuristic} to heuristically argue that FED is unsuited to describe this problem. Based on such arguments, in Sec.~\ref{sectionSystemModel} we propose a physically motivated minimal model to describe the internal quantum dynamics of a levitated nanoparticle in high vacuum. We proceed to exactly solve this model in Sec.~\ref{sectionAnalyticalsol}, matching most of the free parameters with experimentally measurable quantities, and studying the out-of-equilibrium quantum dynamics of the internal degrees of freedom. In particular, the internal energy as function of time is analytically calculated. Finally our conclusions and future perspectives are presented in Sec.~\ref{sectionConclusions}.

\section{How to describe Radiative Thermalization of a Levitated Nanosphere?}\label{SecHeuristic}


In macroscopic bodies, the mechanism for radiative thermalization is well understood as the compensation of an initial energy imbalance via electromagnetic emission and absorption. Indeed, for a particle initially at a higher temperature than the surrounding EM field, the excess of internal energy tends to dissipate into the environment. Such internal energy is held by some internal degrees of freedom (phonons) that cannot interact directly with the EM field. Instead, this interaction is mediated by the fluctuating multipoles generated by the stochastic, phonon-induced motion of the charges forming the body. In particles much smaller than the thermal wavelength ($\lambda_\text{Th} \approx 50\mu\text{m}$ at room temperature) it is customary to take the long-wavelength approximation, so that only the three ($x,y,z$) dipole resonances play a role in the thermalization~\cite{Novotnybook}. It is important to stress that such resonances lie typically in the optical range and thus are very detuned with respect to the thermal energies involved in the process. Hence, they represent a passive channel, allowing the energy exchange between EM field and internal degrees of freedom while always remaining in their ground state~\cite{BiehsBenAbdallah,BenAbdallahPRL2011,ZhuPRB2018,ManjavacasPRB2012,MuletAPL2001}.

While the above physical picture should remain valid for levitated nanoparticles in ultra-high vacuum, we believe that the theoretical approaches based on quasi-equilibrium approximations might become insufficient for an accurate description of thermalization in these systems. On the one hand, for a small particle the  modes describing internal degrees of freedom become highly discretized. For instance, the phononic eigenfrequencies of a spherical particle with radius $R$ and sound velocity $c_s \approx 10^3-10^4$ m$s^{-1}$ can be shown~\cite{Lamb1881} to be the integer multiples of $\omega_\text{phon} \approx \pi c_s/R \approx 2\pi\times 10^{11}$Hz, values which can be comparable to the characteristic frequencies of the EM field, i.e., $k_{\rm B}T_{\rm EM}/\hbar \approx 2\pi \times 10^{12}$Hz at $T_{\rm EM} = 300$K. On the other hand, the combination of levitation and ultra high vacuum should provide an extreme isolation of the nanoparticle from its surroundings, resulting on very narrow phononic linewidths. The combination of high discretization and high isolation of the phononic resonances should have a critical impact on the thermalization dynamics. First, note that, because of the absence of external environments to which the phonons can non-radiatively decay, the internal equilibration times of the nanoparticle might become very large. Indeed, the only mechanism remaining for such equilibration is the internal phonon-phonon interaction which, being such phonons largely detuned with respect to each other, should be much weaker than in bulk. For a small and isolated enough nanoparticle, the internal equilibration times then might become comparable or even larger than the rate at which energy is exchanged with the EM environment, thus violating the quasi-equilibrium assumption. We may thus expect that, in such a regime, the thermalization will occur largely out of equilibrium, making the very concept of temperature ill defined during this process.

A second relevant consequence of the nanoparticle showing highly discrete and narrow phononic resonances is that, except at very high temperatures, only some of the phononic energies are comparable to those of the thermal EM field. Moreover, the coupling between these phonons and the optical resonance will depend on the spatial overlap between the phononic and the dipole mode functions. Since most of these overlaps are negligible by symmetry, only a handful of phononic modes will significantly couple to the EM field. Conversely, most of the phonon modes will be poorly coupled to the EM field. It is then reasonable to expect two intrinsic timescales to arise in the thermalization process. First, the few lowly-detuned modes will thermalize in a short-to-intermediate timescale, during which the largely detuned remaining phonons will remain unperturbed. The latter modes will undergo thermalization only afterwards, at a much larger timescale. In other words, there will be an uneven depletion of the phonon modes, with some of them losing their energy to the EM field much faster than the rest. This is yet a second signature of an out of equilibrium thermalization process where a temperature cannot be assigned to the nanoparticle. 

The two arguments given above, namely long internal thermalization times and uneven depletion of phononic levels, support the breakdown of the quasiequilibrium approach for a sufficiently small nanosphere in high vacuum. Note that
both of such arguments rely on the individual phononic resonances being well resolved in energies, allowing us to roughly estimate the nanoparticle size range at which such effects become appreciable as $\Gamma_\text{phon}/\omega_\text{phon} \ll 1$, where $\Gamma_\text{phon}$ is a typical resonance linewidth. Assuming a spherical particle with radius $R$, and using measured bulk linewidths \cite{TangPRB2011} around $\Gamma_\text{phon} \approx 10^{11}-10^{12}\text{ rad }s^{-1}$ at moderate temperatures, we may estimate that a necessary condition for the local equilibrium assumption to fail is $R\lesssim 10-100$nm, a regime which is accessible for state of the art levitation experiments \cite{Gieseler2012,Asenbaum2013,Jain2016}. Note that the above estimation might be conservative, since phononic linewidths of  isolated nanoparticles are expected to be much smaller than in bulk \cite{ZhangSciAdv2016}.

To summarize, we have given a heuristic argumentation according to which it is reasonable to expect that, at least at short and intermediate timescales, the quasi-equilibrium regime will break down for levitated nanoparticles in high vacuum. In such a case, the thermalization dynamics cannot, by construction, be accurately described by approaches based on \eqnref{eq:dTFD}. This motivates the exploration of new theoretical models able to reproduce the thermalization of such  systems in largely out of equilibrium regimes.

\section{Minimal model}\label{sectionSystemModel}

Motivated by the argumentation above, we devote this section to construct a model for the thermalization of levitated nanoparticles in high vacuum. Contrary to the thermodynamic considerations of FED, our aim is to build a minimal, physically motivated Hamiltonian for the compound NP+EM field system. Such a model should be able to describe the non-equilibrium thermalization dynamics, while at the same time allow us to recover some of the well-known optical and thermal material properties which have been experimentally measured.

The system under consideration is composed of two distinct parts, namely the nanoparticle, fixed at the origin and described by a Hamiltonian $\hat{H}_{\rm NP}$, and the surrounding EM field, with a corresponding Hamiltonian $\hat{H}_{\rm EM}$. These two terms, together with their interaction $\hat{V}$, compose the total Hamiltonian of our system,
\begin{equation}
\hat{H}_{\rm Tot}=\hat{H}_{\rm EM}+\hat{H}_{\rm NP}+\hat{V}.
\label{Hfirst}
\end{equation}
The free EM contribution is well known,
\begin{equation}
\hat{H}_{\rm EM}=\frac{\epsilon_{0}}{2}\int d\mathbf{r}\left(\hat{\mathbf{E}}^{2}+c^{2}\hat{\mathbf{B}}^{2}\right),
\label{HEM}
\end{equation}
where $\hat{\mathbf{E}}$ and $\hat{\mathbf{B}}$ are the transverse electric and magnetic fields, respectively. Note that we do not include the center of mass motion of the NP.

Regarding the NP contribution, we will build its components from physically motivated arguments. First of all, our model must describe the interaction between the NP and the EM field. As discussed above, such interaction can be described, in the long-wavelength approximation, by the dipolar resonances of the NP, such as Mie resonances in dielectrics or plasmon modes in metals \cite{Novotnybook}. Thus, the first component in our NP model will be a set of three dipole modes, one for each orthogonal direction, which we will refer to as \emph{optical degrees of freedom} (ODF). In the usual thermalization scenario, these dipole modes have a negligible excitation probability, since their natural frequency $\Omega$, typically in the optical range, fulfills
 $k_{\rm B}T_{\rm EM} /\hbar \ll  \Omega$ \cite{footnote1}. Moreover, based on the small size of the NP, we neglect any retardation effects on the energy propagation within the body or, equivalently, we assume the time evolution is the same at every point within the NP. More precisely, the NP is regarded as a `lumped system', where the physical quantities are time- but not position dependent \cite{Howell}. 
Both the low excitation probability of the ODF and the `lumped system' assumption allow us to describe these modes as a three-dimensional(3D) point harmonic oscillator, such that the first contribution to the NP Hamiltonian reads
\begin{equation}
\hat{H}_{\FPo}=\frac{\MoPo^{2}}{2\MPo}+\frac{\MPo}{2}\FPo^{2} \QPoDF^{2} = \hbar \Omega \sum_{i=x,y,z} \pare{\adop_i \aop_i + \frac{1}{2}},
\label{HODF}
\end{equation}
where $(\QPoDF)_i= \PoDFc^\text{zpm}  ( \adop_i + \aop_i)$ with $\PoDFc^\text{zpm} = [\hbar/(2 m_\Omega \Omega)]^{1/2}$.
Moreover, since in our model the ODF represent dipole resonances, their interaction with the EM field can be written in the usual dipolar coupling form,
\begin{equation}
\hat{V}=\CAPo~\QPoDF\cdot\hat{\mathbf{E}}\left(0\right) = q \PoDFc^\text{zpm} \sum_{i=x,y,z} \pare{\adop_i + \aop_i} \hat{E_i}\left(0\right) ,
\label{HDip}
\end{equation}
where $q$ is the effective charge that parametrizes the strength of the EM-NP  interaction via the dipole moment of the NP, $q \PoDFc^\text{zpm}$. Note that it is at this point where coupling to other multipoles has been neglected. 


The final components of the model for the NP are the internal degrees of freedom, i.e. the phonons, that thermalize with the EM field through the ODF. As discussed in \secref{SecHeuristic}, the phononic modes of the NP are highly discretized in energies, and only a handful of them are expected to play a significant role. Moreover, in analogy with the excitation of a \emph{collective} dipole mode by the EM field, it is reasonable to expect that, by the same symmetry argument, each dipole mode will mainly couple to one collective phononic excitation. All the remaining modes, being either largely detuned or uncoupled by symmetry considerations, will act as a passive energy reservoir. 
Based on these arguments, we model the internal phonons by splitting them into two groups. First, we introduce three \emph{internal degrees of freedom} (IDF) representing the modes coupling more strongly with the ODF. Due to the `lumped system' assumption, these internal excitations will also be described by a 3D harmonic oscillator, with natural frequency $\omega_\theta$. We therefore include two additional terms in the NP Hamiltonian, namely the energy of such IDF,
\begin{equation}
\hat{H}_{\theta}=\frac{\MoPh^{2}}{2\MPh}+\frac{\MPh}{2}\FPh^{2}\QPhDF^{2}= \hbar \omega_\theta \sum_{i=x,y,z} \pare{\bdop_i \bop_i + \frac{1}{2}},
\label{HTDF}
\end{equation}
and their interaction with the dipole modes, which we assume to be linear,
\begin{equation}
\hat{H}_{\rm Int}=\hbar \CPoPh\frac{\QPoDF}{\PoDFc^\text{zpm}}\cdot\frac{\QPhDF}{\PhDFc^{\text{zpm}}} = \hbar g\!\!\! \sum_{i=x,y,z}\!\!\! \pare{\adop_i + \aop_i}\! \pare{\bdop_i + \bop_i}.
\label{HInt}
\end{equation}
Here $(\QPhDF)_i= \PhDFc^{\text{zpm}}  ( \bdop_i + \bop_i)$ with $\PhDFc^{\text{zpm}} =[\hbar/(2 m_\theta \omega_\theta)]^{1/2}$.
The coupling rate $g$ parametrizes the coupling strength between  the ODF and the IDF.  
In our model, the IDF represent the internal modes with which the EM modes exchange energy more efficiently. 
Consequently, we can assume the frequency of the IDF is not too different from the thermal fluctuations provided by the field, i.e. $\hbar\FPh\approx\mathcal{O}(k_{\rm B}T_{\rm EM})\ll\Omega$\cite{footnote1}. Note that it is also reasonable to expect the coupling rate to be $g \ll \omega_\theta$, since otherwise the IDF-ODF subsystem would be in the ultra-strong coupling regime, which is unphysical in this case.

The second part of the model for the internal phonons of the NP contains the remaining, far detuned internal modes which, as opposed to the IDF, are expected to play a small role in the energy relaxation at moderate time scales. They will be also described by harmonic oscillators, and characterized by a Hamiltonian
\begin{equation}
\hat{H}_{I}=\sum_{n}\left[\frac{\MoBath^{2}}{2\MBath}+\frac{\MBath}{2}\FBath^{2}\QPhBath^{2}\right],
\label{HITB}
\end{equation}
and a coupling to the ODF which we  assume linear,
\begin{equation}\label{HLin}
\hat{H}_{\rm Lin}=\sum_{n}\CPhBath\QPoDF\cdot\QPhBath,
\end{equation}
with a coupling strength given by $\kappa_n$.
Since the impact of these modes on the dynamics is expected to be weak, we will consider them as a passive environment with a fixed temperature, which will be labeled \emph{Internal thermal bath} (ITB). For simplicity, we take the continuum limit of such bath assuming an Ohmic spectral density,
\begin{equation}\label{spectraldensity}
    J_{\gamma_I}(\omega) = \sum_n\frac{\CPhBath^2}{2m_n\omega_n}\delta(\omega-\omega_n) = \frac{2m_\Omega\gamma_I}{\pi}\omega f_c(\omega),
\end{equation}
which is characteristic of phononic environments. In such a way, the ITB is fully described in the minimal model by the rate $\gamma_I$ and a high-energy cutoff function $f_c$~\cite{notecutoffs}. 

\begin{figure}[t] 
	\centering
	\includegraphics[scale=0.30]{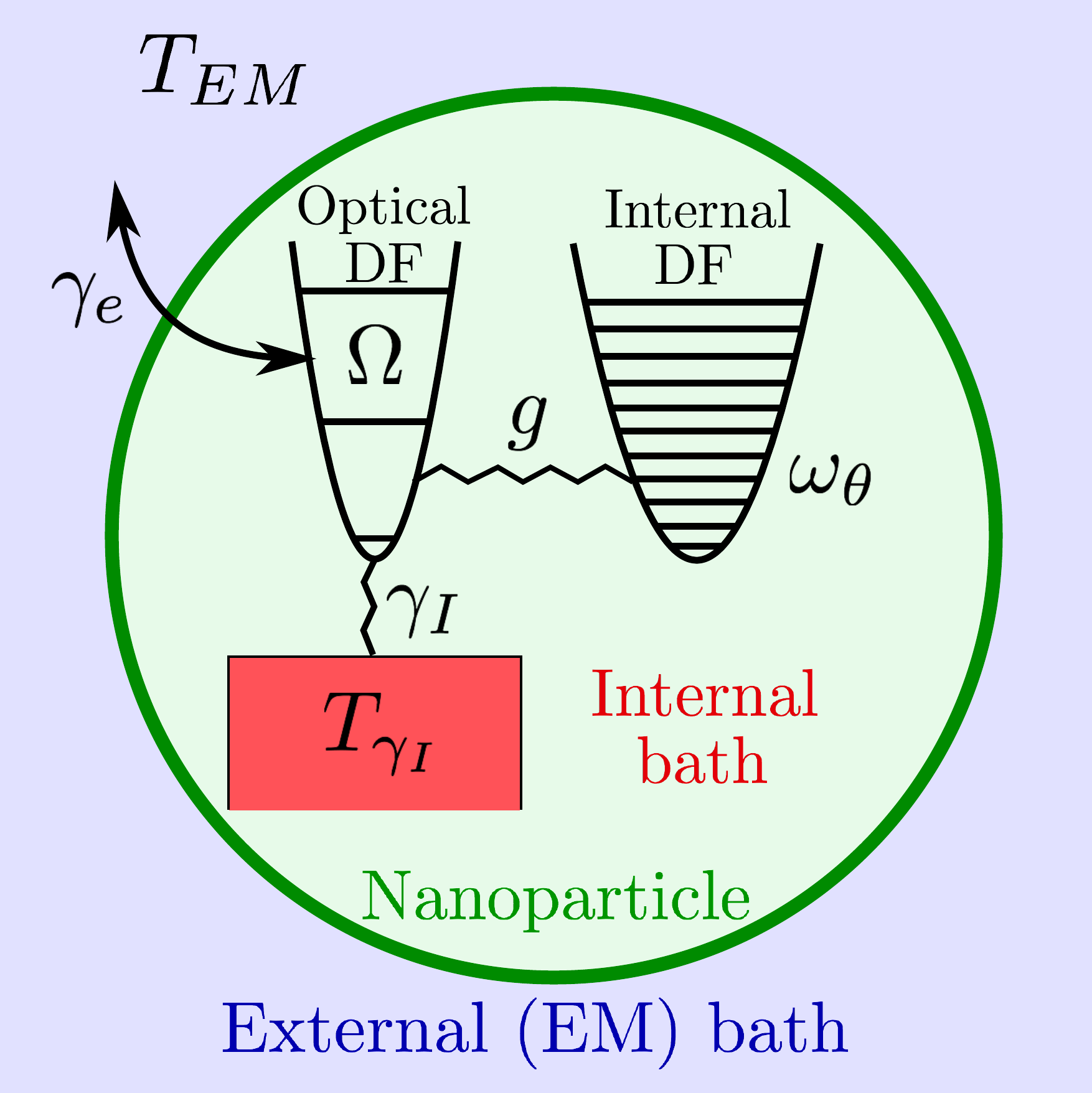}
	\caption{(Color online). Schematic representation of our model for thermalization: the EM field interacts with an internal degree of freedom (IDF) through an optical degree of freedom (ODF), both described as three-dimensional harmonic oscillators. Internal dissipation by other internal modes is included through an internal thermal bath (ITB).}\label{figModel}
\end{figure}

Together with the ODF, the IDF and the ITB form our final model for the nanoparticle, i.e.,
\begin{equation}
    \hat{H}_{\rm NP} = \hat{H}_{\FPo}+\hat{H}_{\theta}+\hat{H}_{\rm Int} + \hat{H}_{I} + \hat{H}_{\rm Lin}.
\end{equation}
A schematic illustration of this minimal model is depicted in \figref{figModel}. It depends on the following free parameters: the frequency of the ODF $\Omega$, the frequency of the IDF $\omega_\theta$, the coupling rate between ODF and IDF $g$, the coupling rate between ODF and ITB $\gamma_I$, and the effective dipole moment of the nanoparticle $q\PoDFc^\text{zpm} \propto q/\sqrt{m_\Omega}$.
Along the following sections we will determine both the physical interpretation of these parameters and their matching with experimental values, summarized in Table \ref{Tableparams}.

We remark that since the temperature $T_{\gamma_I}$ is assumed fixed, the far-detuned phononic modes composing the ITB can never thermalize. However, since, as discussed in Sec.~\ref{SecHeuristic}, such thermalization occurs only at very long times, our model remains valid for the short and intermediate timescales where the out-of-equilibrium dynamics is expected to be relevant. Note that, by construction, our model cannot recover the predictions of FED since it has been built ad-hoc for situations far from quasi-equilibrium. A crossover towards FED could be in principle attained by e.g. substituting the internal thermal bath by a discrete and finite collection of harmonic oscillators. This would allow the bath to dynamically evolve and, consequently, to internally thermalize as well.
Extending the model in this way, however, is not crucial as our current model already allows us to address the short-time dynamics where the quasi-equilibrium approximation is expected to fail, which is the objective of the present work.

As a final comment, note that, because of the interpretation of the TDF and the ITB as the whole set of phononic degrees of freedom, the structure of the two Hamiltonians in Eqs. (\ref{HInt}) and (\ref{HLin}) is the same. Thus, considering the TDF as one of the modes composing the spectral density Eq. (\ref{spectraldensity}), we can integrate the latter in frequencies along the interval $\omega\in[\omega_\theta-\delta,\omega_\theta+\delta]$, where $\delta$ is chosen such that only one phononic peak (the TDF) is included in the integration, i.e. $\delta \lesssim \omega_\theta/2$. This allows us to set an upper bound for the coupling constant $g$ as
\begin{equation}\label{gbound}
    g\lesssim \frac{\omega_{\theta}}{\Omega}\sqrt{\frac{2}{\pi}\frac{\gamma_I}{\Omega}\frac{\delta}{\omega_\theta}}.
\end{equation}
 Given that a lower bound for $\delta$ can only be set by the negligible phonon linewidth, i.e., $\delta \gtrsim \Gamma_{\rm phon} \approx 0$, no lower bound for $g$ can be estimated in the same way.

\begin{table*}[t]
\centering
 \begin{tabular}{ M{2.6cm} || M{2.5cm} | M{2.5cm} | M{3.9cm} | M{2.5cm} | M{2.5cm} } 
   & Dipole resonance frequency & Dipole resonance damping & EM-particle coupling strength & Frequency of internal modes & \vspace{0.35cm} Energy exchange rate \tabularnewline [2.5ex] 
 \hline\hline
 Model parameter & $\Omega$ & $\gamma_I$ & $q^2/m_\Omega$ & $\omega_\theta$ & \vspace{0.2cm}$g$ \tabularnewline  [1ex] 
 \hline
 Physical matching & $\sqrt{\omega_1^2 + \omega^2_{\rm Pl}/3}$ & $\gamma_{\rm D}/4$ & $\epsilon_0 V \omega^2_{\rm Pl}$ & $k_{\rm B}\Theta_{\rm E}/\hbar$ &  \vspace{0.2cm}Free ($g \ll \omega_\theta,\Omega$) \tabularnewline  [1ex] 
 \hline
 Values for Gold & $2\pi\times 1.57 \cdot 10^{15} \rm Hz$ & $10^{-3} \Omega$ & $1.08\cdot10^{-5}\left(R[nm]\right)^3\text{C}^2/\text{Kg}$ & $1.8\cdot10^{-3} \Omega$ &  \vspace{0.2cm} $\lesssim 3.2\cdot 10^{-5} \Omega$ \tabularnewline  [1ex] 
 \hline
 Values for Silica & $2\pi\times 3.39 \cdot 10^{15} \rm Hz$ & $1.8 \cdot 10^{-3} \Omega$ & $5.13\cdot10^{-5}\left(R[nm]\right)^3\text{C}^2/\text{Kg}$ & $2\cdot10^{-3} \Omega$ &  \vspace{0.2cm} $\lesssim 4.8\cdot 10^{-5} \Omega$ \tabularnewline  [1ex] 
 \hline
 \end{tabular}
 \caption{Summary of the relevant free parameters in our model and their physical interpretation. In the second row, the plasma frequency $\omega_{\rm Pl}$, the dielectric damping $\gamma_{\rm D}$, and the Drude-Lorentz resonance $\omega_1$ (equal to zero for metals) refer to
 the generalized permittivity \eqnref{generalpermittivity}, whereas $\Theta_{\rm E}$ represents the Einstein temperature. The bounds for $g$ are calculated from Eq. (\ref{gbound}).}\label{Tableparams}
\end{table*}


\section{Analytical solution and parameter matching}\label{sectionAnalyticalsol}

Once we have set up the Hamiltonian for our model, we are in a position to study the properties of the system. As detailed in Appendices \ref{AppFormalisms} and \ref{MA}, we choose an approach based on the path integral formulation, a method that is especially convenient in our case since all terms in the system Hamiltonian are quadratic \cite{CalzettaHu}. The advantages of this method are twofold: on the one hand, it allows us to determine the full time evolution of the system in a compact way by means of the closed-time-path formalism \cite{CalzettaHu}. Such a procedure results in the exact calculation of the full quantum correlations of our system. On the other hand, it allows for the study of the reduced dynamics of a subsystem (for instance, the EM field alone, or the NP alone) through the so-called Influence Functional, which contains all the information about the external degrees of freedom in the form of relatively simple propagators \cite{CalzettaHu}. It is convenient for this goal to assume an uncorrelated and Gaussian initial state, i.e.
\begin{equation}
\hat{\rho}(0)=\hat{\rho}_{\rm EM}(0)\otimes\hat{\rho}_{\FPo}(0)\otimes\hat{\rho}_{\theta}(0)\otimes\hat{\rho}_{\rm I}(0),
\end{equation}
where the four terms represent the contributions of EM field, ODF, IDF, and ITB, respectively. In this situation, it is possible to \emph{exactly} trace out such external degrees of freedom without any approximation on the system parameters, as opposed to other usual methods such as Born-Markov Master Equations. This includes finite degrees of freedom that cannot be considered as a bath. Given that we make no particular assumption about most of our parameters, and precisely aim at assigning their values through comparison with well-known response functions, our path-integral-based method seems to be a well suited approach.

\subsection{Polarizability of the NP}

Our first goal is to determine the optical response of the NP, i.e., its polarizability according to our model. Such response is calculated by tracing out all the nanoparticle degrees of freedom, i.e. the ODF, the IDF, and the ITB, obtaining a modified equation of motion for the EM field. Specifically, as detailed in Appendix \ref{TOIDF}, we find the following equation of motion for the vector potential in frequency domain,
\begin{equation}
\Big(\nabla\times\nabla\times
-\frac{\omega^{2}}{c^{2}}\left[1+\delta(\mathbf{x})\alpha(\omega)/\epsilon_0\right]\Big)\mathbf{A}(\omega,\mathbf{x})=\mathbf{F}_0.
\label{TransformedEffFieldEq}
\end{equation}
Here, the right-hand side $\mathbf{F}_0$ contains all the terms depending exclusively on the initial conditions of the EM field, which are irrelevant for the calculation of the polarizability. The function $\alpha(\omega)$, for which we have an exact analytical expression in terms of the system propagators, can be identified with the effective polarizability of our model, since it represents the point-like modification to the free EM field evolution. It reads
\begin{equation}
\alpha(\omega)=\frac{\CAPo^2}{\MPo}\left[\FPo^{2}-\omega^{2}-i4\CDamp\omega+\frac{2\FPo\FPh\CPoPh^{2}}{\left(\omega^{2}-\FPh^{2}\right)}\right]^{-1}.
\label{alphamain}
\end{equation}
Note that the above function describes the response of the NP to every frequency of the EM field, and contains multiple resonances associated to the subsystem ODF+IDF.

We are interested in comparing the polarizability obtained in \eqnref{alphamain} to the usual expression for the polarizability of a small NP, namely \eqnref{PolNanoSphere}. The latter expression, however, describes the response to the EM field at frequencies close to the dipole resonance, i.e. in the optical range. On the contrary, the expression extracted from our model, \eqnref{alphamain}, represents such response throughout all the frequency spectrum. Hence, we take the limit of our polarizability for frequencies close to that of the ODF, i.e.
\begin{equation}
\alpha(\omega)\big\vert_{\omega\approx\FPo}\approx\frac{\CAPo^{2}}{\MPo}\left[\FPo^{2}-\omega^{2}-i4\CDamp\omega+\frac{2\FPh\CPoPh^{2}}{\FPo}\right]^{-1},
\label{ApproxPolOptRange}
\end{equation}
where we have used the aforementioned relations $\hbar\omega_\theta \approx \mathcal{O}(k_{\rm B}T_{\rm EM}) \ll \hbar\Omega$\cite{footnote1}.
The above function can be matched to both a dielectric and a metallic particle, through the Drude and Drude-Lorentz permittivity functions respectively \cite{Novotnybook}. To see this, let us define a generalized permittivity function describing both cases as
\begin{equation}\label{generalpermittivity}
\varepsilon_g(\omega)=1+\frac{\omega_{\rm Pl}^{2}}{(\omega_{1}^{2}-\omega^{2}-i\gamma_{\rm D}\omega)},
\end{equation}
where $\omega_1=0$ and $\omega_1 \ne 0 $ for the Drude and the Drude-Lorentz model, respectively. After introducing the above permittivity into the polarizability \eqnref{PolNanoSphere}, a direct comparison with \eqnref{ApproxPolOptRange} yields the following set of necessary identities which, if fulfilled, guarantee that our model reproduces the optical response expected from the NP:
\begin{equation}\label{gammamatching}
\gamma_{\rm D}=4\CDamp,
\end{equation}
\begin{equation}\label{Omegonmatching}
\frac{\omega_{\rm Pl}^{2}}{3} + \omega_1^2=\FPo^{2}+\frac{2\FPh\CPoPh^{2}}{\FPo},
\end{equation}
\begin{equation}\label{qmatching}
\epsilon_{0}V\omega_{\rm Pl}^{2}=\frac{\CAPo^{2}}{\MPo}.
\end{equation}
Since the values of the permittivity $\varepsilon_g(\omega)$ are experimentally tabulated, we will use them together with the above equations to fix the value of three free parameters in our model, namely $\gamma_I$, $\Omega$, and $q/\sqrt{m_\Omega}$ (see Table \ref{Tableparams}).

A few remarks are in place regarding the above parameter matching. First, note that, in the regime of interest $ g \ll \omega_\theta \ll \Omega$, the frequency of the ODF from \eqnref{Omegonmatching} reads $\Omega \approx \sqrt{\omega_1^2+\omega_\text{Pl}^2/3}$, which in general lies, as expected, within the optical range, showing no apparent inconsistencies in our parameter matching. Second, note that it is possible to demonstrate from \eqnref{qmatching} that the ODF is weakly coupled to the EM field, as expected. Indeed, from \eqnref{HDip} we can express the EM-NP coupling rate as $\gamma_e = qx_\Omega^{\text{zpm}} E_{\rm T} / \hbar$, where $E_{\rm T}$ is the thermal electric field defined by $k_{\rm B} T_\text{EM}=  \epsilon_0 \lambda_T^3 E_T^2$ with $\lambda_T = \pi^{2/3} \hbar c / (k_B T_\text{EM})$ being the thermal wavelength. Hence, using \eqnref{qmatching} one obtains
\be
\frac{\gamma_e}{\Omega} =  \pare{\frac{k_B T_\text{EM}}{\hbar \Omega}}^2\sqrt{\frac{ V \omega_\text{Pl}^2 \Omega}{2\pi c^3}}.
\ee 
Since $\omega_\text{Pl} \lesssim \Omega$ and $k_B T_\text{EM} \ll \hbar \Omega $, we can conclude that $\gamma_e/\Omega \ll 1$ for nanoparticles smaller than the optical wavelength $V \Omega^3/c^3 \ll 1$, a condition already assumed in the electric dipole approximation. This weak interaction, which arises naturally from the matching of our parameters to a physical polarizability, will simplify the calculations in the forthcoming sections.


\subsection{Calculation of the internal energy}\label{sectionAsymptSpecificHeat}

The main goal of this article is to study the process of energy thermalization of the NP with the thermal EM field. Because we are interested in exploring non-equilibrium scenarios, it is not convenient (and sometimes not even possible) to assign a temperature to our NP throughout all the time evolution. Thus, we will instead study the evolution of the internal energy of the NP, a well defined observable which, in the quasi-equilibrium limit, is closely related to the temperature (see \eqnref{eq:dTFD}). Note that, in our model, neither the energy of the ODF, which remains unexcited, nor the energy of the ITB, which has constant temperature and is weakly coupled to the ODF, are expected to change appreciably\cite{footnote1}. Thus, the only significant variation of the internal energy will be given by the IDF. Consequently, the observable we identify as internal energy in our model will be $u(t) \equiv \langle\hat{H}_\theta(t)\rangle = \tr [ \hat \rho_\theta(t) \Hop_\theta]$.

In order to calculate the internal energy, we trace out the unrelated degrees of freedom: the EM field (Appendix \ref{TEMF}), the ITB (Appendix \ref{TOIDF}), and the ODF (Appendix \ref{TOODF}), to obtain an effective equation of motion for the IDF (\eqnref{EffEqTDF}). In the process, we assume the initial states of all the system components to be thermal, i.e.,
\begin{equation}
    \hat{\rho}_j(0) \propto e^{-\beta_j \hat{H}_j} \hspace{0.3cm}; \hspace{0.3cm} \left(j = \text{EM},\Omega,\theta,I\right),
\end{equation}
where $\beta_j = (k_{\rm B} T_j)^{-1}$. Moreover, the fact that the ODF and the EM field are weakly coupled to each other, as discussed above, allows us to perform a weak coupling approximation between these two subsystems. This guarantees the absence of runaways (also known as the radiation reaction problem, see e.g. Refs. \cite{Novotnybook,Milonni}) and therefore yields stable solutions (Appendix \ref{TEMF}). The final expression after such approximation simplifies to
\begin{widetext}
\begin{equation}
\begin{split}
 u(t)=&\frac{3\hbar}{4\FPh}\coth\left(\frac{\beta_{\theta}\hbar\FPh}{2}\right)\left(\left[\ddot{\PropPh}
 (t)\right]^{2}+2\FPh^{2}\left[\dot{\PropPh}
 (t)\right]^{2}+\FPh^{4}\left[\PropPh
 (t)\right]^{2}\right)\\
&+\frac{3\hbar}{4\MPh}\int_{0}^{t}d\lambda d\lambda'\Big[\dot{\PropPh}
(t-\lambda)\KerNoise_{\rm Env}(\lambda,\lambda')\dot{\PropPh}
(t-\lambda')+\FPh^{2}\PropPh
(t-\lambda)\KerNoise_{\rm Env}(\lambda,\lambda')\PropPh
(t-\lambda')\Big].
\label{EnergyNormalizedTime}
\end{split}
\end{equation}
\end{widetext}
Written in this way, the above expression has a clear interpretation. The first line directly represents the effective evolution associated to the initial state of the IDF. Specifically, it accounts for the value of the energy at $t=0$ and, additionally, for the relaxation dynamics of the IDF due to the dissipation induced by the remaining subsystems (the ODF, the ITB, and the EM field), through the retarded propagator $\PropPh
(t)$.
On the other hand, the second line accounts for the fluctuations that such subsystems induce on the IDF, through the noise kernel $\KerNoise_{\rm Env}(t,t')$. 
According to the effective dynamics we obtain for the IDF, the fluctuations are provided by each part of the system and funnelled through the retarded propagator, leading to terms of the form $\PropPh
\KerNoise_{\rm Env} \PropPh
$. Since the first line of \eqnref{EnergyNormalizedTime} accounts exclusively for the relaxation dynamics, it is clear that the contribution at infinite time (i.e. the steady state) will be provided by the second line, that is, the contribution of the fluctuations. A more detailed explanation of all the functions involved in \eqnref{EnergyNormalizedTime}, together with their analytical expressions, can be found in Appendix \ref{GFCF}.

Before studying the full thermalization dynamics described by \eqnref{EnergyNormalizedTime}, we discuss asymptotic limits that allow us to  match the parameters of the minimal model.

\subsection{Asymptotics of thermalization and specific heat}

Although the components of the internal energy \eqnref{EnergyNormalizedTime} are obtained analytically, calculating its value as a function of time requires a numerical integration \cite{notecutoffs}. However, the structure of the propagators and kernels allows to simplify the equation in some limits. For instance, it is possible to show that at short times, $t \to 0$, the internal energy reads
\begin{equation}\label{shorttime}
    u(t\approx 0)\approx u_{0} + \frac{\omega_\theta g^2}{\Omega}u_{\Omega 0}t^{2}\left[1-\frac{1}{2}\left(4\CDamp+\frac{\FPo^{2}}{\omega_{q}}\right)t\right].
\end{equation}
Here, $u_0 = (3\hbar\omega_\theta/2)\coth\left(\beta_\theta\hbar\omega_\theta/2\right)$ and $u_{\Omega 0 } = (3\hbar\Omega/2)\coth\left(\beta_\Omega\hbar\Omega/2\right)$ represent the initial energies of the IDF and the ODF, respectively, while the frequency $\omega_q = 6\pi m_\Omega c^3 \epsilon_0/q^2$ is defined for convenience. In the above equation, the first term corresponds to the initial energy, whereas the second term, which contains quadratic but not linear corrections, describes the evolution of the initial thermal state of the IDF caused by its interaction with
the remaining degrees of freedom. At sufficiently short times such that the quadratic terms dominate, only the ODF contribute to the dynamics, indicating a certain degree of retardation in the influence of the continuous environments (EM and ITB). In this regime, the energy shows an increasing behavior, stemming from the coupling between the IDF-ODF being turned on at $t=0$ in our calculations. However, at later times, when the cubic term becomes relevant, the continuous environments start affecting the dynamics, introducing a negative counterterm that results in the expected decrease in energy. 
The time at which the increasing tendency is reverted, i.e., the maximum of \eqnref{shorttime} is $t_{\rm max}=(4/3)[4\CDamp+(\FPo^2/\omega_{q})]^{-1}\approx 4\omega_{q}/(3\FPo^{2}) \approx 0.05$ fs
for a gold nanoparticle with $R=50$nm. This value is irrelevant for the thermalization timescales we are interested in. However, note that having a Hamiltonian allows us to calculate the evolution at arbitrarily short timescales,
something that cannot be done by quasi-equilibrium models such as FED. 

\begin{figure}[h!] 
	\centering
	\includegraphics[width=\linewidth]{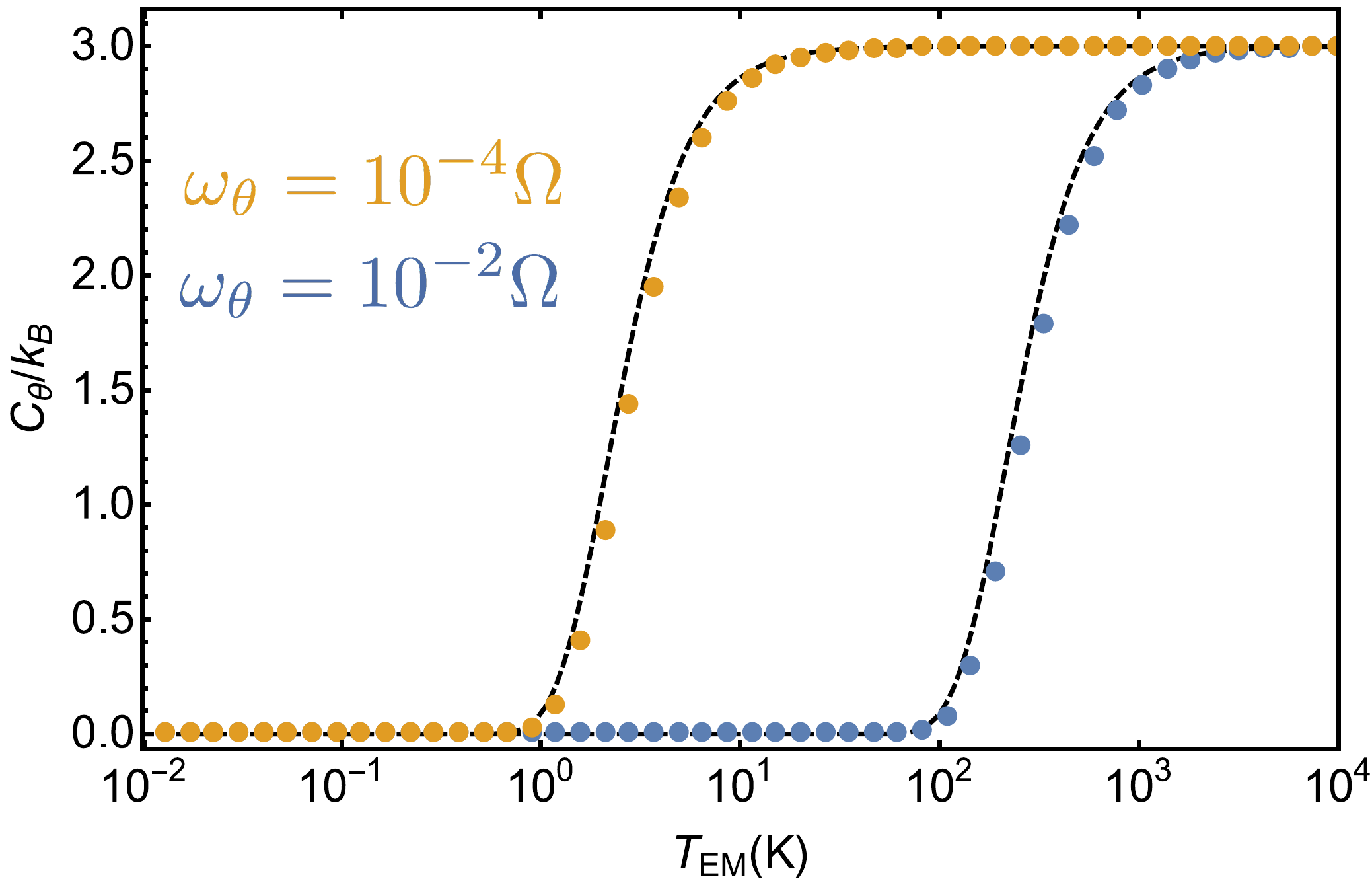}
	\caption{(Color online). Specific heat as predicted by \eqnref{specificheat} for different frequencies $\omega_\theta$ of the IDF. The dashed black lines show the Einstein specific heat of a three-dimensional harmonic oscillator with natural frequency $\omega_\theta$. This plot corresponds to a gold NP with a radius $R = 50$nm, and a coupling $g = 10^{-9}\Omega$.}\label{figspecificheat}
\end{figure}

Let us now focus on the opposite limit, namely the long-time behavior of \eqnref{EnergyNormalizedTime}. Note that, even though we are taking $t\to \infty$ in our expression, the resulting asymptotic behavior represents only the long time limit within the intermediate timescales where our model is valid. It can be shown that in such limit the NP thermalizes with the environment, as the internal energy tends to the following constant (see Appendix \ref{LBTDFE}),
\begin{equation}\label{Hinfinity}
\begin{split} 
    u_\infty=& u(t\to \infty) =\frac{3\hbar \FPh\CPoPh^{2}\FPo}{2\pi}\int_{0}^{\infty}\!\!\!\!d\omega K(\omega) \times
    \\
    &
    \times\!\bigg\{\!
    \frac{\FPo^{2}}{\omega_q}\coth\!\left[\frac{\beta_{\rm EM}}{2}\hbar\omega\right] + 
     4\CDamp\coth\!\left[\frac{\beta_{\CDamp}}{2}\hbar\omega\right]\!\!\bigg\}.
\end{split}
\end{equation}
The function $K(\omega)$ can be constructed from the retarded propagators and is given by
\begin{equation}
\begin{split}
    K\!(\omega)\! = \!\frac{m_\Omega^2}{q^4}\frac{\omega(\omega_\theta^2+\omega^2)}{(\omega_\theta^2-\omega^2)^2}\left\vert\frac{\alpha(\omega)}{1 - i\omega\Omega^2\alpha(\omega)/(6\pi\epsilon_0c^3)}\right\vert^2\!\!\!.
\end{split}
\end{equation}
In the expression above, the term inside the modulus squared corresponds to the effective polarizability employed in the literature to include the corrections of radiation reaction \cite{Novotnybook}. Note that the long time limit has two contributions, associated to the two baths (EM and ITB) which continuously provide fluctuations to the IDF. 

In analogy with the polarizability, the long-time limit of the internal energy, \eqnref{Hinfinity}, provides us with a tool to recover the thermal response of the NP, since in the long time limit we expect such NP to be in, or very close to, thermal equilibrium. This allows us to assign thermodynamic properties to our NP, specifically a heat capacity, and use it to fix the value of a fourth free parameter in our model. In order to define the heat capacity we note that, based on the optical matching Eqs.~(\ref{gammamatching}-\ref{qmatching}), it can be shown that $\gamma_I \ll \Omega^2/\omega_q$. Thus, in the second line of \eqnref{Hinfinity}, the contribution of the ITB is negligible, and the whole thermal dynamics will be mainly governed by the free EM field. To define the specific heat, we consider a small variation in the temperature of the EM field, which according to \eqnref{Hinfinity} will induce a corresponding variation in the energy of the IDF. As mentioned above, this energy exchange can be assigned to the whole nanoparticle since, on the one hand, the ODF remains unexcited and, on the other hand, the ITB is a passive bath with no absorption. In other words, we can interpret the variation of $u(t)$ with respect to $T_{\rm EM}$ as the heat exchanged between NP and EM field upon changing the temperature of the latter. Thus, at long times, we can define the heat capacity of the NP in terms of our internal energy in the usual way,
\begin{equation}\label{specificheat}
\HCT(\beta_{\rm EM})\equiv-k_{\rm B}\beta_{\rm EM}^{2}\frac{\partial u_\infty}{\partial\beta_{\rm EM}}.
\end{equation}

The specific heat extracted from our model is plotted in \figref{figspecificheat} for a gold NP with radius $R=50$nm. Note that, once the polarizability has been matched with empirical data, only two free parameters remain, namely the frequency of the IDF, $\omega_\theta$, and its coupling to the ODF, $g$. Regarding the specific heat, however, the dependence with the latter only becomes relevant for very large values of the coupling constant, $g\gtrsim \omega_\theta$. Since, as mentioned above, we focus on the physical regime where $g \ll \omega_\theta$, our expression for the specific heat effectively depends only on the parameter $\omega_\theta$. This suggests that, being the IDF a 3D harmonic oscillator, \eqnref{specificheat} could be matched with the usual expression given by the Einstein model \cite{Ashcroft},
\begin{equation}\label{Einstein}
    C_\text{E}(\beta_{\rm EM}) = 3k_{\rm B}\left(\hbar\omega_\theta\beta_{\rm EM}\right)^2\frac{e^{\hbar\omega_\theta\beta_{\rm EM}}}{\left(e^{\hbar\omega_\theta\beta_{\rm EM}}-1\right)^2}.
\end{equation}
The above Einstein formula turns out to be in excellent agreement with our results, as illustrated by the dashed lines in \figref{figspecificheat}. On the one hand, this certifies that, as expected, the NP is in thermal equilibrium in the long time limit. On the other hand, the matching between \eqnref{specificheat} and \eqnref{Einstein} allows us to fix the value of the free parameter $\omega_\theta$ through the phenomenological Einstein temperature $\Theta_{\rm E}= \hbar\omega_\theta/k_{\rm B}$, which can be determined experimentally. For most materials, this parameter lies in the range $\Theta_{\rm E}\approx 100-2000$K, which implies $\omega_\theta \approx 2\pi \times\left(10^{12}-10^{13}\right)$ Hz. Note that our matching is consistent, as such frequencies fulfill the physically motivated assumption $k_{\rm B}T_{\rm EM}\approx\omega_\theta \ll \Omega$ used in previous sections \cite{footnote1}. Additionally, the values obtained for $\omega_\theta$ are compatible with the phononic frequencies of nanometric particles, i.e. $\omega_\text{phon} \approx 2\pi \times\left(n \cdot 10^{13}/R\text{[nm]}\right)$ Hz ($n=1,2,3...N$, with $N$ being the number of atoms composing the particle).

\subsection{Full  thermalization dynamics}\label{sectionDynamicsComparisonFED}

\begin{figure}[h!] 
	\centering
	\includegraphics[width=\linewidth]{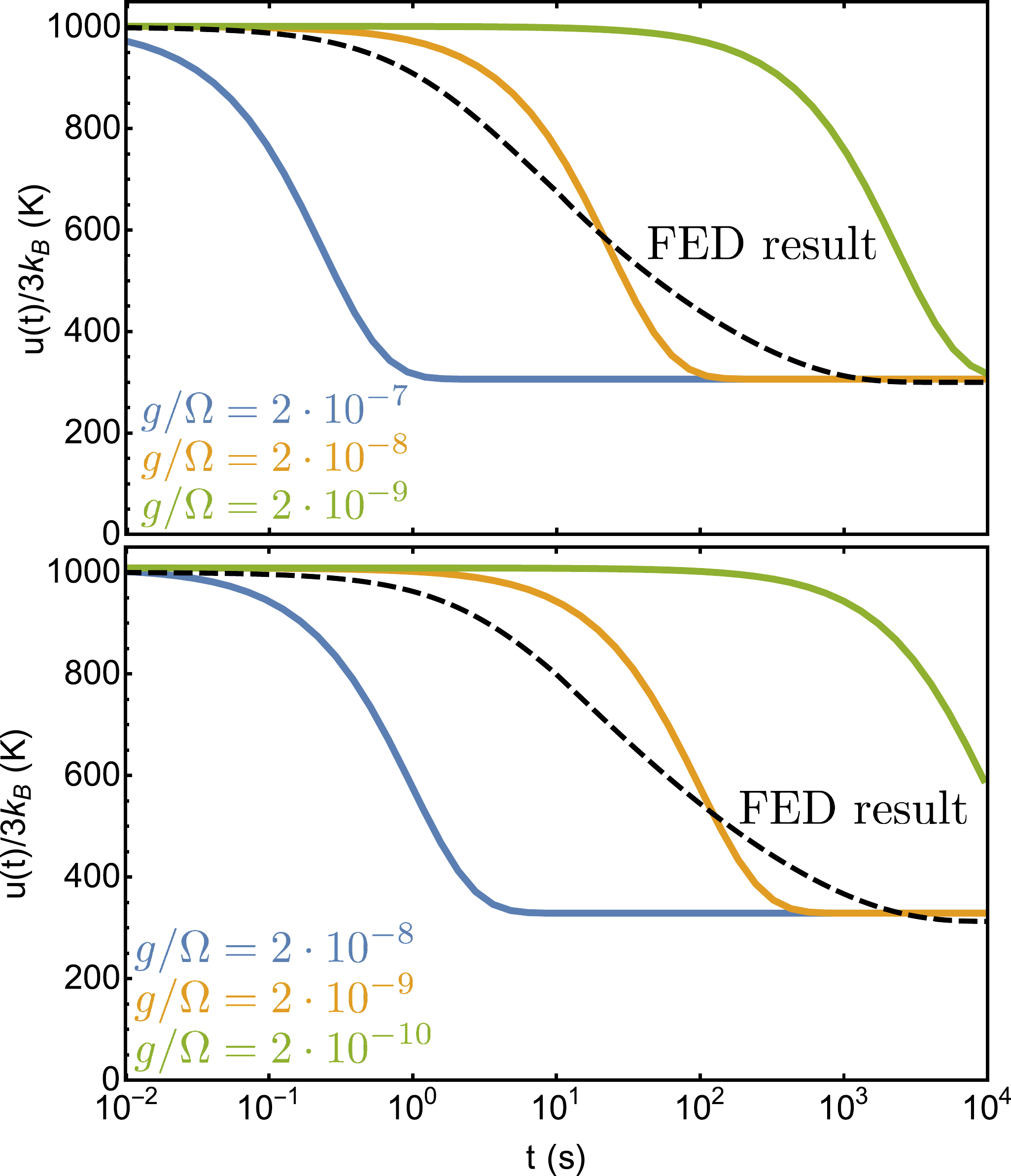}
	\caption{(Color online). Time evolution of the internal energy as given by \eqnref{EnergyNormalizedTime}, for a gold (upper panel) and silica (lower panel) NP with radius $R=50$nm and initial temperature $T=1000$K, interacting with an EM field at $T_{\rm EM}=300$K. The colored lines show three different values of the coupling rate between ODF and IDF. The dashed line illustrates the FED result, \eqnref{eq:dTFD}.}\label{figTimeEvol}
\end{figure} 

Once we have matched the optical and thermal response of our NP, we can study the full time evolution, \eqnref{EnergyNormalizedTime}, as a function of the only remaining free parameter, $g$. Since such parameter describes the coupling between the IDF and the remaining degrees of freedom, it will be responsible for the timescale of the thermalization. At this point, in order to clearly describe this thermalization process, it is convenient to define a notion of `temperature' for the relevant degree of freedom.
 Specifically, since the magnitude $u(t)$ represents the internal energy of a set of three degenerate harmonic oscillators (the IDF), it is reasonable to consider the magnitude $u(t)/3k_{\rm B}$ as the closest definition of temperature in our system. Indeed, it is possible to show that $u(t)/3k_{\rm B}\to T$ for a NP in thermal equilibrium in the limit $T \gg \Theta_{\rm E}$. An example of such tendency can be easily observed in the thermal (and thus equilibrium) initial state of the NP, whose internal energy is given by \eqnref{shorttime}.
However, let us remark that $u(t)/3k_{\rm B}$ is only meaningful in energetic terms as, in general, our system is not in thermal equilibrium at any time during the evolution and, consequently, a temperature \emph{cannot} be assigned to the entire NP.

\begin{figure}[h!] 
	\centering
	\includegraphics[width=\linewidth]{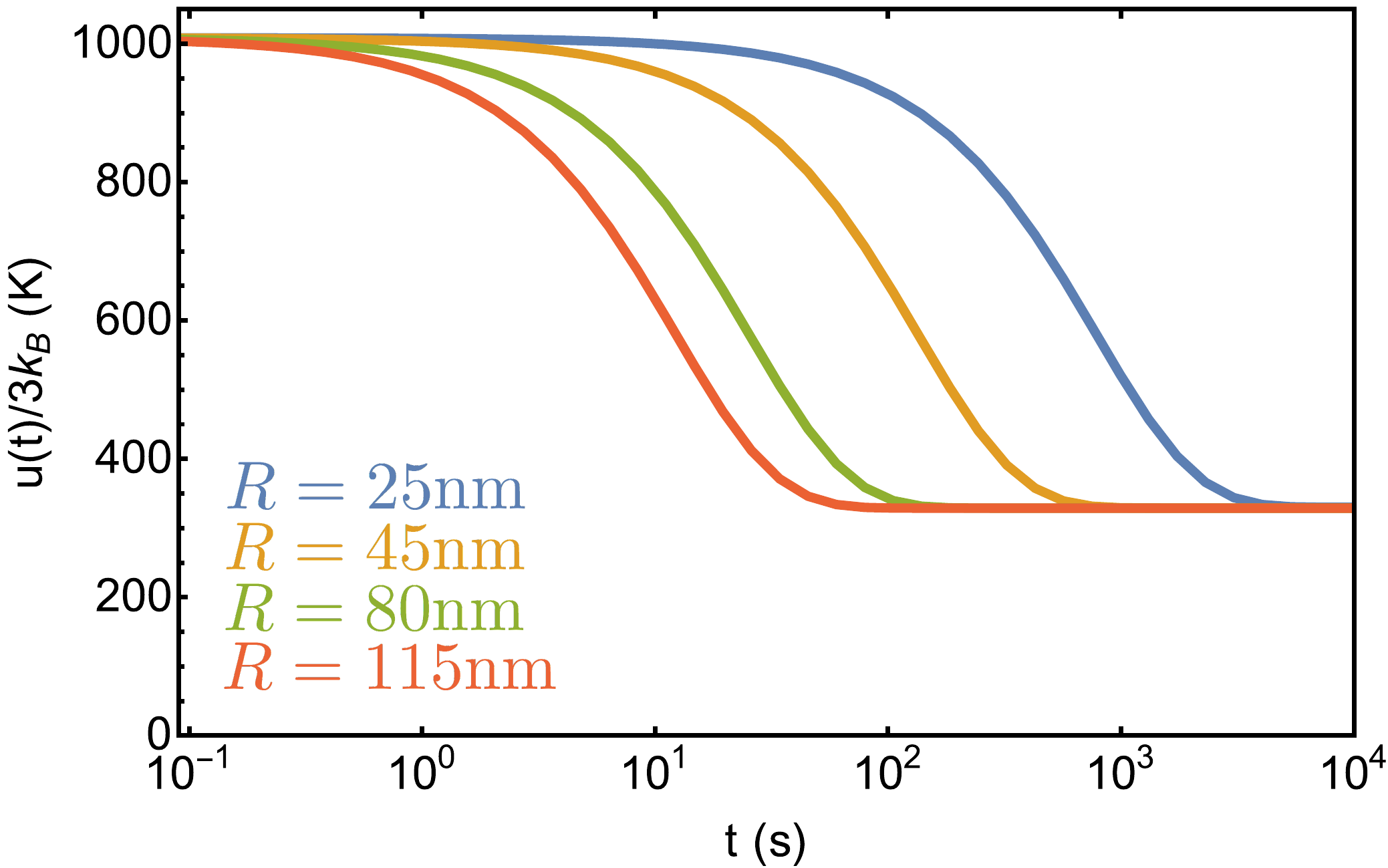}
	\caption{(Color online). Time evolution of the internal energy for Silica and different NS radii according to the parameters given in Table \ref{Tableparams}. Similar results are found for gold (by rescaling the couplings $g$ as shown in Fig. \ref{figTimeEvol}) .}\label{figdifferentradii}
\end{figure}

The full time evolution of the internal energy for both gold and silica is shown in \figref{figTimeEvol}, assuming an initial NP temperature $T_\Omega = T_\theta = T_{\gamma_I} = 1000$K and an EM field temperature $T_{\rm EM}=300$ K. Interestingly, all three colored curves in each panel, corresponding to different values of the remaining free parameter of our model, $g$, share a common shape, the effect of modifying $g$ being only a global horizontal shift proportional to $ g^2$. This fact is consistent with our definition of the internal energy as the energy of the IDF alone which, in order to equilibrate with the EM environment, must funnel its extra energy to the ODF at a rate given by $g$. In other words, we can physically interpret $g$ as the energy exchange rate of the NP. Note that the value of $g$ yielding a thermalization timescale consistent with that of \figref{fig1} is $g \approx 10^{-8}\Omega$ for gold and $g \approx 10^{-9}\Omega$ for silica. Moreover, the same qualitative behavior remains when varying the radius of the NP, as shown in Fig. \ref{figdifferentradii}. As expected, a larger radius results in a larger polarizability (i.e., a larger EM field-ODF coupling), thus inducing a faster thermalization.

The full internal energy relaxation dynamics calculated with our model can be compared with the results of macroscopic electrodynamics, namely FED. As shown by the dashed lines in \figref{figTimeEvol}, the results obtained with our model differ not only quantitatively but qualitatively with FED. Indeed, whereas the energy in our model decays following a multi-exponential behavior, FED predicts an approximately polynomial decay, faster at short times and slower close to the final state. This discrepancy, which as discussed in Sec. \ref{SecHeuristic} was to be expected in the way we constructed our model, could be resolved by experiments capable of measuring \figref{figTimeEvol}. This would also determine the free parameter of our model, $g$, which critically characterizes the relaxation timescale of the nanoparticle.
Indeed, experiments attempting at measuring the internal temperature of an optically levitated nanosphere have been recently reported~\cite{HebestreitPRA2018}.

\section{Conclusions}\label{sectionConclusions}



In this work, we have addressed the problem of the thermalization of a levitated nanoparticle in high-vacuum, by developing a physically motivated minimal model describing the interaction between the nanoparticle and the thermal electromagnetic field. First, we give a detailed description of the physical aspects of the problem, which leads us to infer that the commonly used assumption of quasi-equilibrium might not hold for the entire nanoparticle. Specifically, on the one hand, the extreme isolation of the internal phonon modes should lead to very high internal thermalization times whereas, on the other hand, the uneven coupling between such phonons and the external environment is expected to induce a faster depletion of some of these modes.
Motivated by this, we have built a physically motivated minimal model able to account for these out-of-equilibrium processes. 
Then, we have exactly solved the dynamics of our model using path integral techniques and the influence functional method. 
This has allowed us to reproduce both the optical and the thermal response of a nanoparticle in terms of the polarizability and the specific heat, respectively. 
Building on the above results, we have studied the thermalization dynamics of the nanoparticle. We have shown that it occurs largely out of equilibrium, differing from the predictions of approaches based on the quasi-equilibrium assumption, such as fluctuation electrodynamics.
Although, to our knowledge, no experiments have yet reported the thermalization dynamics of a levitated nanoparticle in high vacuum, promising recent works \cite{HebestreitPRA2018} suggest that reaching this out-of-equilibrium regime lies well within experimental reach.

Our minimal model attempts at theoretically studying the extreme physical regimes of ultra high-vacuum levitated nanomatter. It hints at the possibility that conventionally accepted approximations and regimes, both in condensed matter and in light-matter interaction physics, are unsuited to describe this scenario. We believe that, in the following years, a deeper theoretical study of these systems, combined with their experimental availability, will unravel plenty of exotic and unexplored physical phenomena.

\begin{acknowledgments}
We acknowledge support by the Austrian Federal Ministry of Science, Research, and Economy (BMWFW). CGB acknowledges funding from the European Union’s Horizon 2020 research and innovation programme under the Marie Sk\l{}odowska-Curie grant agreement No H2020-MSCA-IF-2017-796725. We further thank the hospitality of the Centro de Ciencias de Benasque Pedro Pascual, where this manuscript was completed.

AERL and CGB contributed equally to this work.
\end{acknowledgments}


\clearpage

\appendix
\section{Basics on functional methods: Influence functional and closed-time-path formalisms}\label{AppFormalisms}

Here and in the following appendices, we set $\hbar\equiv 1$ for simplicity, and use Einstein summation convention over repeated indices. 
In this appendix, we present a basic introduction to the path integral methods used in the article. In order to address the quantum dynamics of time-dependent, out of equilibrium systems, we will combine two formalisms. On the one hand, the closed-time-path method, which allows us to study the time evolution of quantum systems in terms of functional integrals. On the other hand, the influence functional method, suitable for addressing the effective dynamics of a system in presence of another one considered as an environment.

Our starting point is to express the matrix elements of the time evolution operator, $\langle x_{t}|\hat{U}(t)|x_{0}\rangle$, in terms of a path integral. Here, $x_t$ represents an arbitrary complete set of dynamical variables at time $t$. As detailed in Ref.  \cite{CalzettaHu}, the above matrix element is determined by discretizing the time interval $(0,t)$ into infinitesimal time steps which, in the continuum limit, become an infinite set of zero-length intervals. This allows us to write
\begin{equation}
\langle x_{t}|\hat{U}(t)|x_{0}\rangle=\int^{x(t)=x_{t}}_{x(0)=x_{0}}Dx~e^{iS},
\label{UIntegral}
\end{equation}
where $\int Dx$ represents a functional integral over all the trajectories $x(t)$ passing through the initial and final points, and $S$ is the classical action expressed in terms of $x$ and its temporal derivatives. As detailed in Ref.  \cite{CalzettaHu}, the above equation has been obtained by explicitly carrying out the integration in the momentum variables assuming a quadratic kinetic term. Thus, it
is generally valid for Hamiltonians with quadratic dependence on the momentum operator. 


\subsection{Closed-Time-Path integrals}

The fundamental representation of the time evolution operator in \eqnref{UIntegral} allows us to describe the dynamics of the density matrix and, consequently, of any expectation value. Indeed, it is straightforward to write the matrix elements of the density operator in the Heisenberg picture as
\begin{equation}
\begin{split}
\rho(x,x'&,t)=\langle x|\hat{U}(t)\hat{\rho}\hat{U}^{\dag}(t)|x'\rangle
\\
&
=\int dx(0)\int dx'(0)~\rho(x(0),x'(0),0)
\\
&
\times\int_{x(0)}^{x(t)=x}Dx\int_{x'(0)}^{x'(t)=x'}Dx'e^{i(S[x]-S[x'])}.
\end{split}
\label{RhoIntegrales}
\end{equation}
This expression shows the density matrix elements written in terms of a path integral along two \emph{independent} histories, i.e. the paths $x$ and $x'$. In general, however, we will not be interested in the whole density matrix, but on the correlations between the operators $\hat{X}$ at different times, i.e. $G_{12}(\tau,\tau')\equiv\langle \hat{X}(\tau')\hat{X}(\tau)\rangle$. 
In order to calculate such correlators in a compact and elegant way, we will use the closed-time-path (CTP) formalism.

Our first step is to define a more general difference action than in \eqnref{RhoIntegrales}, by introducing two artificial linear sources, $J(t)$ and $J'(t)$, for the two paths, i.e.
\begin{equation}\label{AccionFuncionalGeneratriz}
    \begin{split}
        S_{\rm CTP}[x,J;x',J']&=S[x]-S[x']+
        \\
        &
        +\!\int_0^t\!\!d\lambda\left[J(\lambda)x(\lambda)-J'(\lambda)x'(\lambda)\right].
    \end{split}
\end{equation}
We can now use the above CTP action to define the \emph{generating functional}, i.e. a functional of the two sources $J$ and $J'$, which, as we detail below, will contain all the information about the correlations we seek. This generating functional is defined as
\begin{equation}\label{FuncionalGeneratriz}
    \begin{split}
        Z[J,J']&=\int dx\int dx(0)\int dx'(0)\rho(x(0),x'(0),0)
        \\
        &\times\int^{x(t)=x}_{x(0)}\!\!\!Dx\int^{x'(t)=x}_{x'(0)}\!\!\!Dx'e^{iS_{\rm CTP}[x,J;x',J']},
    \end{split}
\end{equation}
and is closely related to the evolution of the density matrix (notice the similarity with \eqnref{RhoIntegrales}). However, it is important to remark that, unlike the density operator, the two path integrals in \eqnref{FuncionalGeneratriz} are not independent due to the tracing over the final time variable, i.e., the integral over $x$ together with the \emph{common} upper limit for the two path integrals. The fact that the two histories are connected through their final point allows the interpretation of the double integral as a single path integral along a unique history defined in a closed-time-path, $x(0)\to x \to x'(0)$.

According to the definition of the generating functional, it is straightforward to demonstrate  \cite{CalzettaHu} that every correlation can be calculated as a functional derivative of $Z[J,J']$, i.e., 
\begin{equation}
    \begin{split}
        G_{12}(\tau,&\tau')  = 
        \\
        &
        =\!\left[\frac{1}{i}\frac{\delta}{\delta J(\tau)}\right]\!\!\left[-\frac{1}{i}\frac{\delta}{\delta J'(\tau')}\right]\!Z[J,J']\Big|_{J,J'=0}.
    \end{split}\label{G12derivatives}
\end{equation}
Thus, the problem of finding correlations is reduced to the calculation of the generating functional. Importantly, in cases where the momentum can be defined as $p=m\dot{x}$, any other correlation such as, for instance, that of two momentum operators, is also deducible from the generating functional \cite{CalzettaHu}. Note that although the expression \ref{G12derivatives} is independent of the temporal ordering between $\tau$ and $\tau'$, it is possible to also obtain time-ordered correlations from $Z$.


Aside from calculating correlations, the CTP action defined in \eqnref{AccionFuncionalGeneratriz} can be used to derive the equation of motion for the physical degree of freedom $x$, through the prescription
\begin{equation}
\frac{\delta S_{\rm CTP}[x;x']}{\delta x}\Big|_{x=x',J=J'=0}=0.
\label{EcMovCTP}
\end{equation}
As we will see in the next section, the equation of motion resulting from \eqnref{EcMovCTP} is real and causal, even for open systems with non-unitary evolutions. It is important to remark also that the equation deduced in this way can eventually include dissipation but not noise. In fact, such an equation can be interpreted as averaged over all the noise realizations. In the case of open dynamics, however, it is possible to obtain a Langevin-type equation (including noise) by including in the imaginary part of the CTP effective action a stochastic source with a given distribution as a variable in the functional integrations. The resulting equation can be related to the Heisenberg equation of motion for the corresponding operator, albeit with a different interpretation (See Ref.  \cite{CalzettaRouraVerdaguer} for more details). In the following section, we will detail how the effective dynamics of open systems can be addressed using the path integral formalism.

\subsection{Influence functional}

Let us consider a system $S$ represented by a degree of freedom $x$, in interaction with an environment $E$ described by the set of degrees of freedom $q=\{q_{n}\}$. The hamiltonian of the total system is $\hat{H}=\hat{H}_{\rm S}+\hat{H}_{\rm E}+\hat{H}_{\rm int}$, with
\begin{equation}
\hat{H}_{\rm S}=\frac{\hat{p}^{2}}{2M}+V(\hat{x})~~~;~~~\hat{H}_{\rm int}=V_{\rm int}(\hat{x},\hat{q}).
\label{HamiltonianoS-Int}
\end{equation}
For simplicity we will assume that $\hat{H}_{\rm E}$ is a quadratic function of the momentum operators $\hat{p}_n$. 
In analogy with the previous section, the quantum state of the compound $S+E$ system is given by its density matrix $\hat{\rho}(t)$, which evolves in a unitary fashion from its initial state $\hat{\rho}(0)$, i.e.  $\hat{\rho}(t)=e^{-it\hat{H}}\hat{\rho}(0)~e^{it\hat{H}}$. Therefore, we can write this evolution in terms of a path integral as shown in \eqnref{RhoIntegrales}, using the classical action $S[x,q]=S_{\rm S}[x]+S_{\rm E}[q]+S_{\rm int}[x,q]$, as
\begin{widetext}
\begin{equation}
\begin{split}
\rho(x~q,x'q',t)&=\langle x~q,t|\hat{\rho}|x'q',t\rangle
=\int dx_{i}dq_{i}\int dx'_{i}dq'_{i}\langle x~q,t|x_{i}q_{i},0\rangle\langle x_{i}q_{i},0|\hat{\rho}|x'_{i}q'_{i},0\rangle\langle x'_{i}q'_{i},0|x'q',t\rangle
\\
&
=\int dx_{i}dq_{i}\int dx'_{i}dq'_{i}\int_{x_{i}}^{x} Dx\int_{q_{i}}^{q} Dq~ e^{iS[x,q]}~\rho(x_{i}q_{i},x'_{i}q'_{i},0)\int_{x'_{i}}^{x'}Dx'\int_{q'_{i}}^{q}Dq'~e^{-iS[x',q']}
\\
&
\equiv \int dx_{i}dq_{i}\int dx'_{i}dq'_{i}~K(x~q,x'q,t|x_{i}q_{i},x'_{i}q'_{i},0)~\rho(x_{i}q_{i},x'_{i}q'_{i},0).
\end{split}
\label{RhoInfluence}
\end{equation}
\end{widetext}
In the last step, we have defined $K$ as the propagator of the total system. This function contains all the information of the internal system dynamics and is the object we aim at computing.

In the usual open system scenario, one is interested in the dynamics of the system $S$, the detailed dynamics of the environment $E$ being unnecessary. More specifically, we are interested in the expectation values of operators with the form $\hat{A}\otimes\mathbb{I}_{E}$, where $\hat{A}$ is an operator acting only on the Hilbert space of system $S$. Such expectation values can be calculated from the reduced density matrix of $S$, $\hat{\rho}_{r}$, obtained by tracing over the environmental degrees of freedom, i.e., $\hat{\rho}_{r}={\rm Tr}_{q}(\hat{\rho})$. The matrix elements of the reduced density operator can thus be written as
\begin{equation}
\rho_{r}(x,x',t)={\rm Tr}_{q}(\hat{\rho})=\int_{-\infty}^{+\infty}dq~\rho(x~q,x'q,t).
\label{RhoReducida}
\end{equation}
Assuming that the $S-E$ interaction is turned on at $t=0$, and that the initial state is uncorrelated, $
\rho(x_{i}q_{i},x'_{i}q_{i}',0)=\rho_{\rm S}(x_{i},x'_{i},0)\otimes\rho_{\rm E}(q_{i},q'_{i},0)
$, we can easily take the trace over the total density matrix elements in \eqnref{RhoInfluence} to obtain
\begin{equation}
\begin{split}
\rho_{r}(x,x',t)&=\int dx_{i}dx'_{i}
\\
&
\times
K_{r}(x,x',t|x_{i},x'_{i},0)\rho_{r}(x_{i},x'_{i},0),
\end{split}
\label{RhoReducidaEvol}
\end{equation}
which is written in terms of a new propagator for the reduced density matrix,
\begin{equation}
\begin{split}
K_{r}(x,x'&,t|x_{i},x'_{i},0)\equiv
\\
&
\equiv\int_{x_{i}}^{x}Dx\int_{x'_{i}}^{x'}Dx'~e^{i(S[x]-S[x'])}F[x,x'].
\end{split}
\label{OpEvolReduc}
\end{equation}
Note that \eqnref{RhoReducidaEvol} has the form of the of the evolution of the density matrix of a closed system. However, the propagator \eqnref{OpEvolReduc} does not only contain the free evolution phase $S[x]-S[x']$, but also the \emph{Feynman-Vernon influence functional},
\begin{equation}
    \begin{split}
        &F[x,x'] 
        =\int dqdq_{i}dq'_{i}~\rho_{\rm E}(q_{i},q'_{i},0)
        \\
        &
        \times\!\!\!\int_{q_{i}}^{q}\!\!\!Dqe^{i(S_{\rm E}[q]+S_{\rm int}[x,q])}\!\!\!
        \int_{q'_{i}}^{q}\!\!\!Dq'e^{-i(S_{\rm E}[q']+S_{\rm int}[x',q'])}.
    \end{split}
    \label{FuncionalFeynmanVernon}
\end{equation}
This influence functional accounts for the effect of the environment on the system dynamics, being $F[x,x']=1$ in the case of a closed system, i.e. when $\hat{H}_\text{int}=0$.
Usually, it is convenient to define the \emph{Influence action} $S_{\rm IF}$ through 
\begin{equation}\label{influenceaction}
    F[x,x'] = e^{iS_{\rm IF}[x,x',t]}.
\end{equation}
This allows us to interpret the influence of the environment as an extra term in the action that couples the paths on the functional integral \ref{OpEvolReduc}, $x$ and $x'$. Thus, in open quantum systems, the calculation of the generating functional $Z$ is more involved, since it requires calculating the influence functional and then performing the coupled double path integral.

The influence functional formalism is very suitable for treating open quantum systems. Indeed, it leads to Langevin-type equations for the effective dynamics of the systems, where the effect of the environment appears through both dissipation and noise. In the next section, we analyze the specific case of linear couplings, which is a cornerstone across many areas involving the study of dissipative dynamics.



\subsection{Linear couplings}

Here, we will detail the expression for the calculation of the generating functional $Z$ in a more specific scenario .
Let us consider that the environment action is quadratic in its variables $q$, and assume the initial state is Gaussian. Furthermore, we will assume the interaction term of the action to be linear, 
$S_{\rm int}=\int dt~x^{a}(t)~q_{a}(t)$. Here, the scripts denote summation over the different CTP branches, i.e. $x^a \in (x,x')$, while the coordinates $q_a$ can include a coupling constant. Under all these assumptions, the influence functional in \eqnref{FuncionalFeynmanVernon} can be interpreted as the functional Fourier transform of a Gaussian functional with variables $q$ and $q'$. As a result, the influence action is necessarily a quadratic form of $x$ and $x'$, i.e., $S_{\rm IF}=(1/2)\int dtdt'~x^{a}(t)\mathbb{M}_{ab}(t,t')x^{b}(t)$. Here, the matrix elements are by definition equal to
\begin{eqnarray}
\mathbb{M}_{ab}(t,t')=-i
\frac{\delta^{2}}{\delta x^{a}(t)\delta x^{b}(t')}e^{iS_{\rm IF}[x^{a},t_{\rm f}]}\Big|_{x^{a}=0},
\label{MatrizInfluencia-CTP}
\end{eqnarray}
where $t_{\rm f}$ is the final time of the considered evolution. 

Importantly, thanks to the assumptions undertaken above, the matrix elements of $\mathbb{M}_{ab}(t,t')$ can also be analytically calculated by directly taking the variations of the influence action in \eqnref{influenceaction}. Specifically, it can be shown that such elements depend on the expected values of the quantum operators associated to the dynamical variable $q$, i.e.,
\begin{equation}
\mathbb{M}_{ab}(t,t')=i
\left(\begin{array}{clrr} %
\left\langle T[\hat{q}(t)\hat{q}(t')]\right\rangle & -\left\langle \hat{q}(t')\hat{q}(t)\right\rangle \\
-\left\langle \hat{q}(t)\hat{q}(t')\right\rangle & \left\langle \tilde{T}[\hat{q}(t)\hat{q}(t')]\right\rangle \\
\end{array}\right).
\label{MatrizInfluencia-CTPMatriz}
\end{equation}
Here, $T$ and $\tilde{T}$ represent, respectively, the time ordering and reverse time ordering operator, and the expected values are taken without considering the interaction with the system, i.e. evolving only with $\hat{H}_{\rm E}$. Using the above result, we can explicitly write the influence action in terms of the system variables as
\begin{widetext}
\begin{equation}
\begin{split}
    S_{\rm IF}=\frac{i}{2}\int dtdt'
    \Big[\left\langle T[\hat{q}(t)\hat{q}(t')]\right\rangle x(t)x(t')&-\left\langle \hat{q}(t')\hat{q}(t)\right\rangle x(t)x'(t')
    \\
    &
    -\left\langle \hat{q}(t)\hat{q}(t')\right\rangle x'(t)x(t')+\left\langle \tilde{T}[\hat{q}(t)\hat{q}(t')]\right\rangle x'(t)x'(t')\Big].
    \end{split}
    \label{AccionInfluenciaPropagadores}
\end{equation}
A more convenient expression in order to determine the generating functional can be obtained by changing to the sum and difference variables, $x+x'$ and $x-x'$, and using the relations between the different correlation functions:
\begin{equation}
    S_{\rm IF}=\int dtdt'\Big[(x-x')(t)D(t,t')(x+x')(t')
    +\frac{i}{2}(x-x')(t)N(t,t')(x-x')(t')\Big].
\label{AccionInfluenciaNucleosDyN}
\end{equation}
\end{widetext}
The first term above is related to the dissipation in $x$, and is governed by the so-called dissipation kernel $D(t,t')$. Such dissipation kernel corresponds to a retarded propagator,
\begin{eqnarray}
D(t,t')=\frac{i}{2}\left\langle [\hat{q}(t),\hat{q}(t')]\right\rangle\theta(t-t').
\label{NucleoDisipacion}
\end{eqnarray}
On the other hand, the second term is associated to the fluctuations induced in $x$ by the environment. These fluctuations are characterized by the noise kernel, which corresponds to a Hadamard propagator
\begin{eqnarray}
N(t,t')=\frac{1}{2}\left\langle \{\hat{q}(t),\hat{q}(t')\}\right\rangle.
\label{NucleoRuido}
\end{eqnarray}
It is important to note that, given the hermiticity of the involved operators, both kernels are real, while the dissipation kernel is also causal. 

The influence action given in \eqnref{AccionInfluenciaNucleosDyN} allows us to directly calculate the generating functional for a system in contact with an environment. This calculation is performed by simply introducing the influence function $S_{\rm IF}$ as an extra term of the CTP action $S_{\rm CTP}$ defined in \eqnref{AccionFuncionalGeneratriz} and \ref{FuncionalGeneratriz}. In this case, however, the integrations are not so straightforward as in the case of a closed system, since the two paths $x$ and $x'$  are coupled by the influence action term. Nevertheless, the computation of $Z$ can be carried out also in this case, as proven in Ref. \cite{CalzettaRouraVerdaguer} for the action used in this paper (and generalized to fields in Ref. \cite{CTPGauge}). The procedure requires to obtain the equations of motion for the system degrees of freedom using \eqnref{EcMovCTP}. From such equations, we extract their associated propagators $G_{\rm Ret}$, with initial conditions set to $G_{\rm Ret}(0)=0,\dot{G}_{\rm Ret}(0)=1$. Then, it is possible to write the paths in terms of such Green functions and obtain a compact expression for the generating functional:
\begin{equation}
\begin{split}
Z[J,J']&=\left\langle e^{-iJ_-\ast X_{0}}\right\rangle_{X_{i},p_{i}}e^{-iJ_-\ast G_{\rm Ret}\ast J_+}\\
&\times e^{-\frac{1}{2}J_-\ast G_{\rm Ret}\ast N\ast(J_-\ast G_{\rm Ret})^{T}},
\end{split}
\label{generalgeneratingfunctional}
\end{equation}
with $J_- = J'-J$ and $J_+ = (J'+J)/2$.
Here, we have defined the operation $A\ast B\equiv\int dtA(t)B(t)$. Note that such operation can be concatenated by adding integrations, e.g. $J\ast G_{\rm Ret} \ast J'\equiv \int dt\int dt' J(t)G_{\rm Ret}(t,t')J'(t')$. Additionally, the expected value is taken using the initial Wigner function of our state, $W(X,p,0)$, as $\left\langle ...\right\rangle_{X_{i},p_{i}}\equiv\int_{-\infty}^{+\infty}dX_{i}\int_{-\infty}^{+\infty}dp_{i}...W(X_{i},p_{i},0)$. Finally, $X_{0}$ is the homogeneous solution of the effective equation of motion obtained through \eqnref{EcMovCTP}, written in terms of the initial conditions $X_{i},p_{i}$:
\begin{equation}
X_{0}(t)=\dot{G}_{\rm Ret}(t)~X_{i}+G_{\rm Ret}(t)~\frac{p_{i}}{M}.
\end{equation}

It is important to remark that in many situations, including our model, there is not a single environment but an ensemble of layered environments which have to be traced out sequentially. For instance, in the main text, we obtain an effective equation of motion for the EM field by tracing first the IDF and the ITB to get an effective evolution for the ODF, and afterwards continue by tracing such ODF in the final step. In these situations, the influence action always has a similar structure to that of \eqnref{AccionInfluenciaNucleosDyN}. However, after performing more than one trace, the corresponding dissipation and noise kernels become more involved: on the one hand, the dissipation kernel becomes an effective retarded propagator including the retarded propagators of the external systems already traced out. On the other hand, the noise kernel is not anymore a Hadamard propagator, but a more involved object including both the noise and the dissipation kernels of the traced out environments. 
Regardless of this difference, the generating functional can always be calculated by successive application of \eqnref{generalgeneratingfunctional}.

\section{The model actions}\label{MA}

This appendix is devoted to give the actions that are associated to the hamiltonians of \eqnref{Hfirst}-\ref{HLin}. Note that our formalism is based on \eqnref{UIntegral}, where we assumed a quadratic dependence of the system Hamiltonian in all the canonical momenta in order to integrate such variables out. As a result, the classical actions appearing in the functional integrals depend on the variables of the degrees of freedom ($x$ in the previous section) and their time derivatives, but not on their momenta. Hence, we have to explicitly write the classical actions in this form.

From now one and throughout all the appendices, greek subscripts will denote both temporal and spatial coordinates, whereas latin subscripts will indicate spatial coordinates only. Under this convention, the spacetime metric will be $\eta_{\mu\nu}=(1,-1,-1,-1)$, and the four-potential which describes the EM field will be $A^{\mu}$. 
Analogously, three-dimensional vectors such as the electric field and other quantum degrees of freedom will be written either with a latin subscript $E_j$ or, or equivalently, in vector form as $\mathbf{E}$.

The total action of the system must contain seven terms, corresponding to the free actions of the four components of our system (EM field, ITB, ODF, IDF) and the three interaction terms (see \figref{figModel} for illustration). Using the same notation as in the main text, we thus have
\begin{widetext}
\begin{equation}\label{Totalactionsystem}
    S\left[A^{\mu},\PoDF,\PhDF,\{\PhBath\}\right]
    =S_{\rm EM}[A^{\mu}]+S_{I}\left[\{\PhBath\}\right]+S_{\FPo}[\PoDF] + S_{\theta}[\PhDF]
    +
    S_{\rm Dip}\left[A^{\mu},\PoDF\right]+S_{\rm Int}\left[\PoDF,\PhDF\right]+S_{\rm Lin}\left[\PhDF,\{\PhBath\}\right].
\end{equation}
\end{widetext}
The first four terms correspond to the free actions, and are given by
\begin{equation}
\begin{split}
    S_{\rm EM}[A^{\mu}]&=-\frac{\epsilon_{0}}{4}\int_{t_{\rm in}}^{t_{\rm f}} d^{4}xF_{\mu\nu}F^{\mu\nu}
    \\
    &
    =\frac{\epsilon_{0}}{2}\int_{t_{\rm in}}^{t_{\rm f}}d^{4}x\left(\mathbf{E}^{2}-c^{2}\mathbf{B}^{2}\right),
\end{split}
\end{equation}
\begin{equation}
S_{\FPo}[\PoDF]=\int^{t_{\rm f}}_{t_{\rm in}}d\lambda\frac{\MPo}{2}\left[\dot{\mathbf{x}}_\Omega^{2}(\lambda)-\FPo^{2}\PoDF^{2}(\lambda)\right],
\end{equation}
\begin{equation}
S_{\theta}[\PhDF]=\int^{t_{\rm f}}_{t_{\rm in}}d\lambda\frac{\MPh}{2}\left[\dot{\mathbf{x}}_\theta^{2}(\lambda)-\FPh^{2}\PhDF^{2}(\lambda)\right],
\end{equation}
\begin{equation}
S_{\rm I}\left[\{\PhBath\}\right]=\sum_{n}\int^{t_{\rm f}}_{t_{\rm in}}d\lambda\frac{\MBath}{2}\left[\dot{\mathbf{x}}_n^{2}(\lambda)-\FBath^{2}\PhBath^{2}(\lambda)\right],
\end{equation}
where $F_{\mu\nu}=\partial_{\mu}A_{\nu}-\partial_{\nu}A_{\mu}$ is the EM tensor (with $\partial_{\mu}=\left(\frac{1}{c}\frac{\partial}{\partial t},\nabla\right)$), and $\lambda$ is a time variable. On the other hand, the last three terms in \eqnref{Totalactionsystem} represent the interaction actions, given by
\begin{equation}
S_{\rm Dip}\left[A^{\mu},\PoDF\right]=\CAPo\int^{t_{\rm f}}_{t_{\rm in}}d\lambda~\PoDF(\lambda)\cdot\mathbf{E}\left(z^{\mu}_{0}\right),
\label{Sdip}
\end{equation}
\begin{equation}
S_{\rm Int}\left[\PoDF,\PhDF\right]=2\sqrt{\MPo\MPh\FPo\FPh}\CPoPh\!\int^{t_{\rm f}}_{t_{\rm in}}\!\!d\lambda\PoDF(\lambda)\!\cdot\!\PhDF(\lambda),
\end{equation}
\begin{equation}
S_{\rm Lin}\left[\PhDF,\{\PhBath\}\right]=\sum_{n}\CPhBath\int_{t_{\rm in}}^{t_{\rm f}}d\lambda~\PoDF(\lambda)\cdot\PhBath(\lambda),
\label{IntPhDFPhBath}
\end{equation}
where for the sake of clarity we indicate by $\mathbf{z}_0$ the position of the particle (taken at the origin in the main text), such that $z^{\mu}_{0}=\left(\lambda,\mathbf{z}_{0}\right)$.
In the calculations, we will rewrite the EM-ODF interaction \eqnref{Sdip} in the more convenient, Lorentz-invariant form
\begin{equation}
S_{\rm Dip}\left[A^{\mu},\PoDF\right]\equiv S_{\rm Dip}\left[J_{\mu},A^{\mu}\right]=\int d^{4}xJ_{\mu}(x)A^{\mu}(x),
\end{equation}
with an effective current given by
\begin{equation}\label{defcurrents}
\begin{split}
J_{\mu}(x)=-\CAPo\int^{t_{\rm f}}_{t_{\rm in}}&d\lambda
\PoDFc^{j}(\lambda)
\\
&\times
\left(\eta_{j\mu}\partial_{0}+c\eta_{0\mu}\partial_{j}\right)\delta\left(x^{\alpha}-z^{\alpha}_{0}\right).
\end{split}
\end{equation}









\section{Tracing out the internal degrees of freedom}\label{TOIDF}

\begin{table*}[t]
\centering
 \begin{tabular}{ M{2.1cm} || M{2.2cm} | M{2.2cm} | M{2.2cm} | M{2.3cm} | M{2.2cm} | M{2.2cm} } 
   & 
	\includegraphics[width=0.8\linewidth]{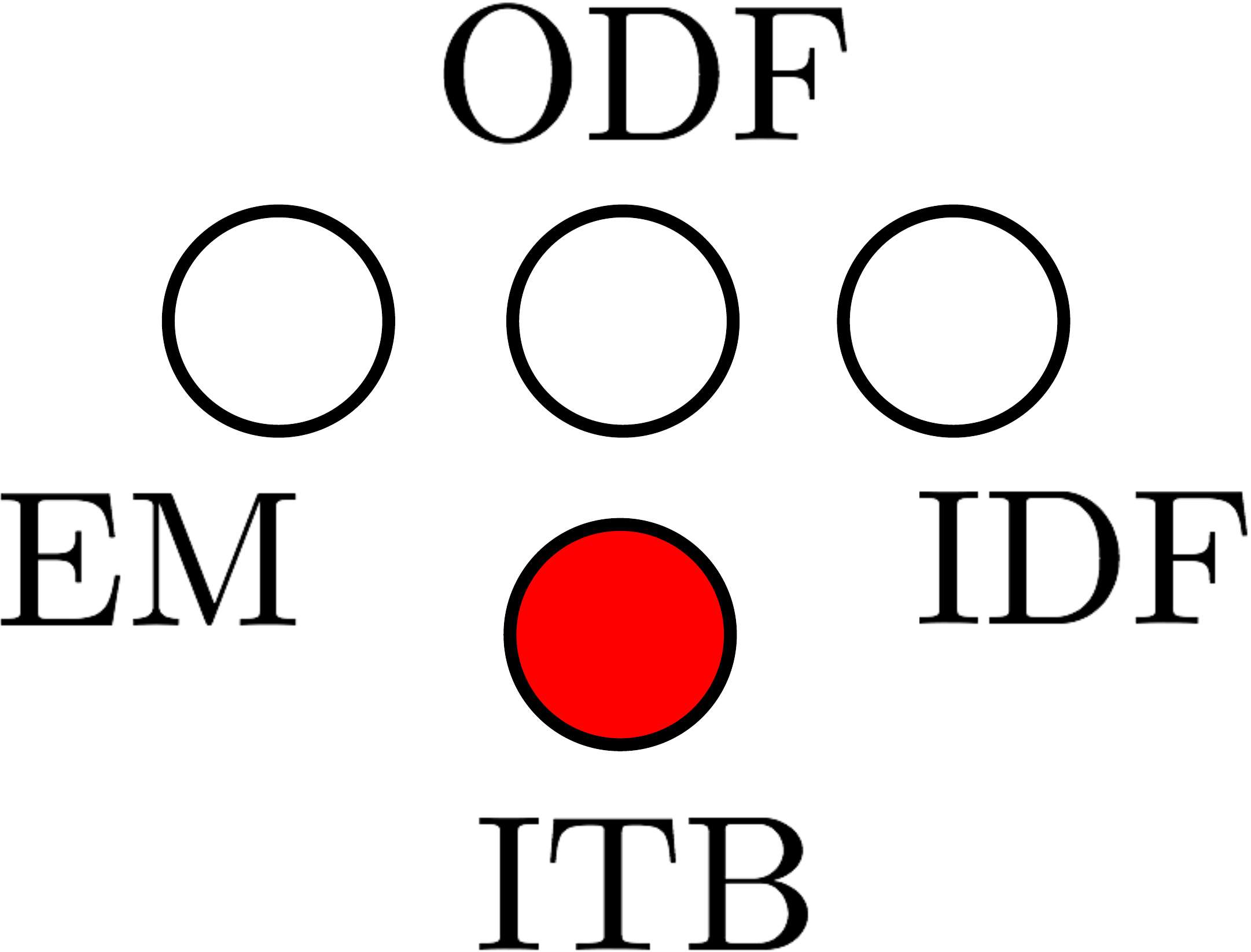}
   & 
   \includegraphics[width=0.8\linewidth]{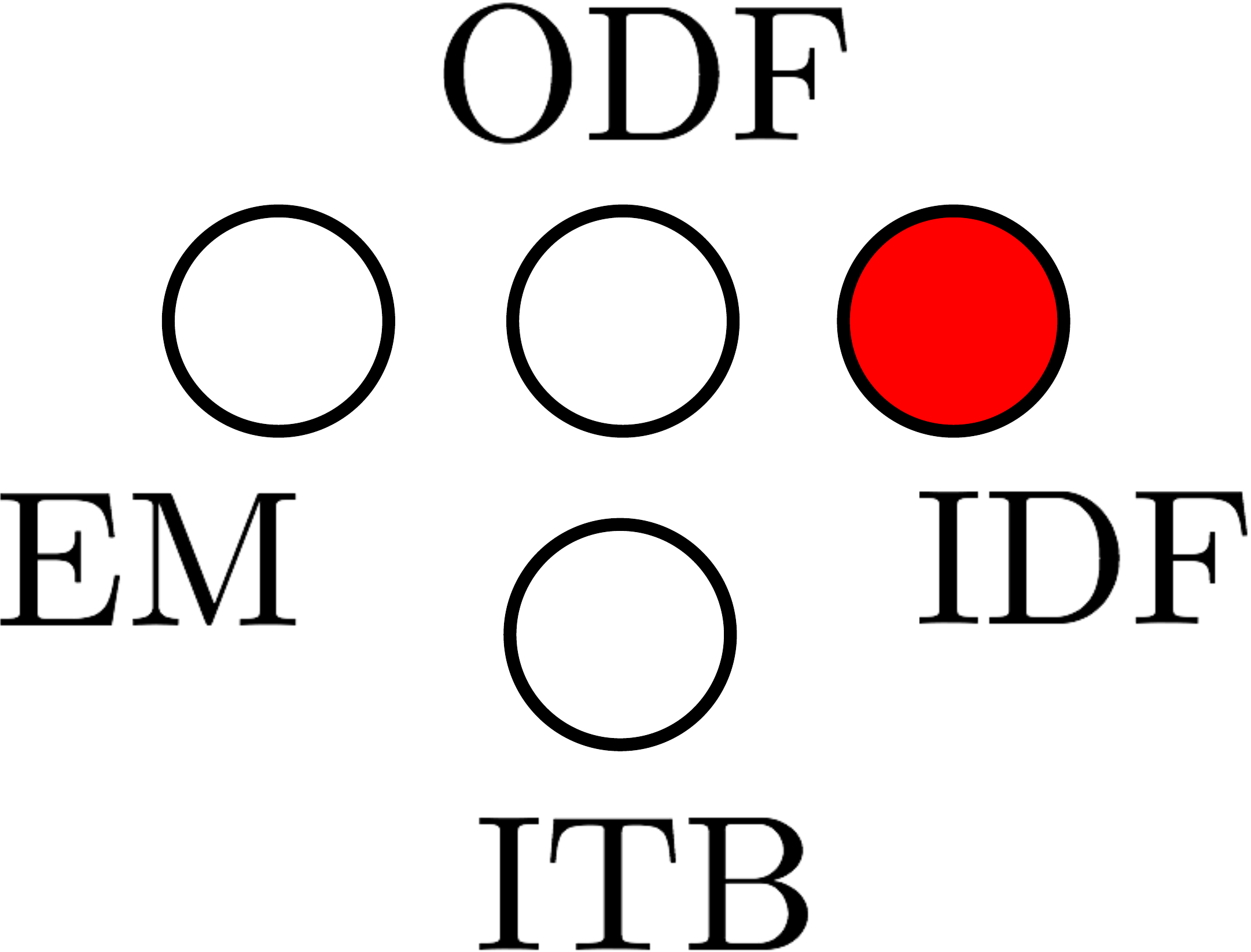}
   & 
   \includegraphics[width=0.8\linewidth]{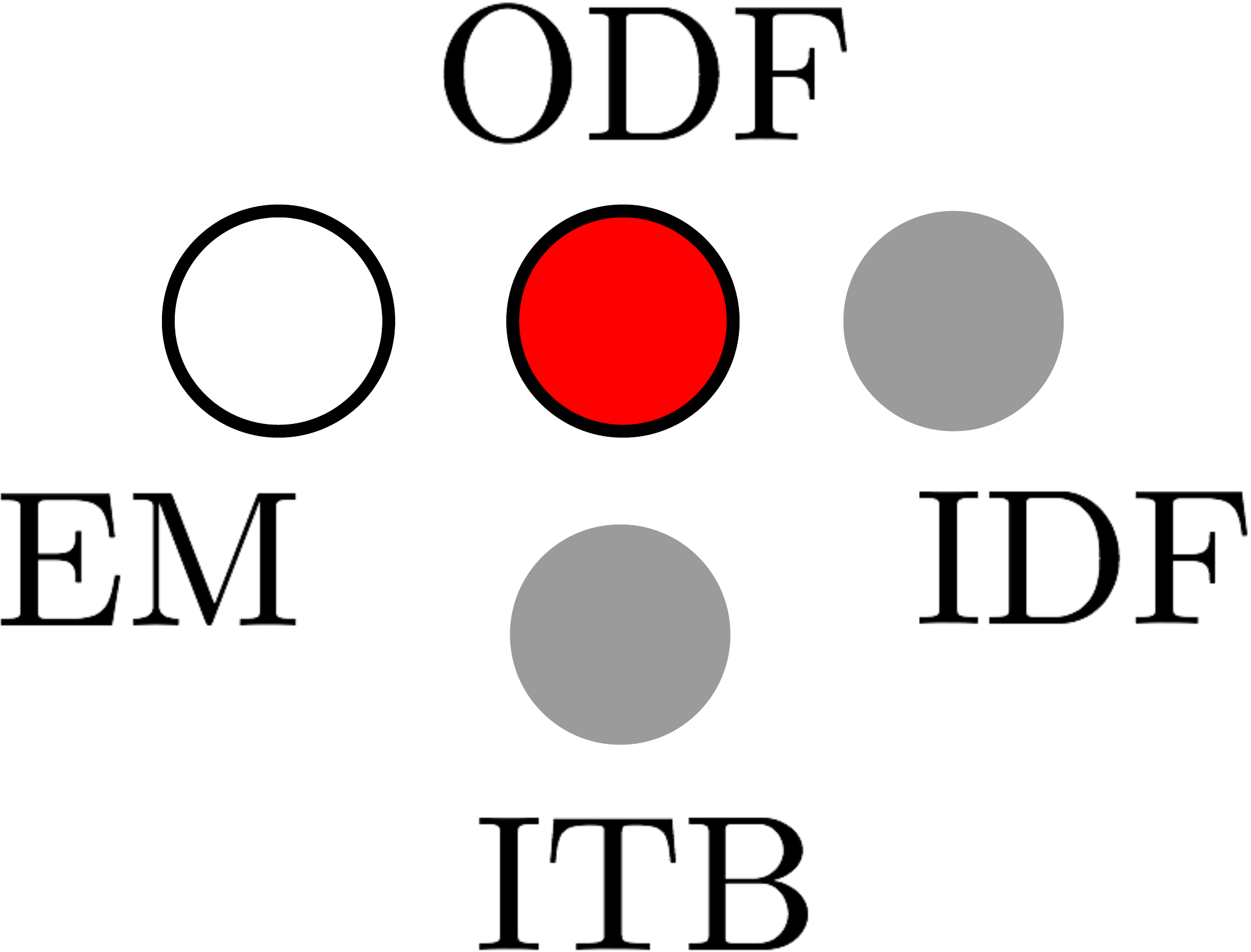}
   & 
   \includegraphics[width=0.8\linewidth]{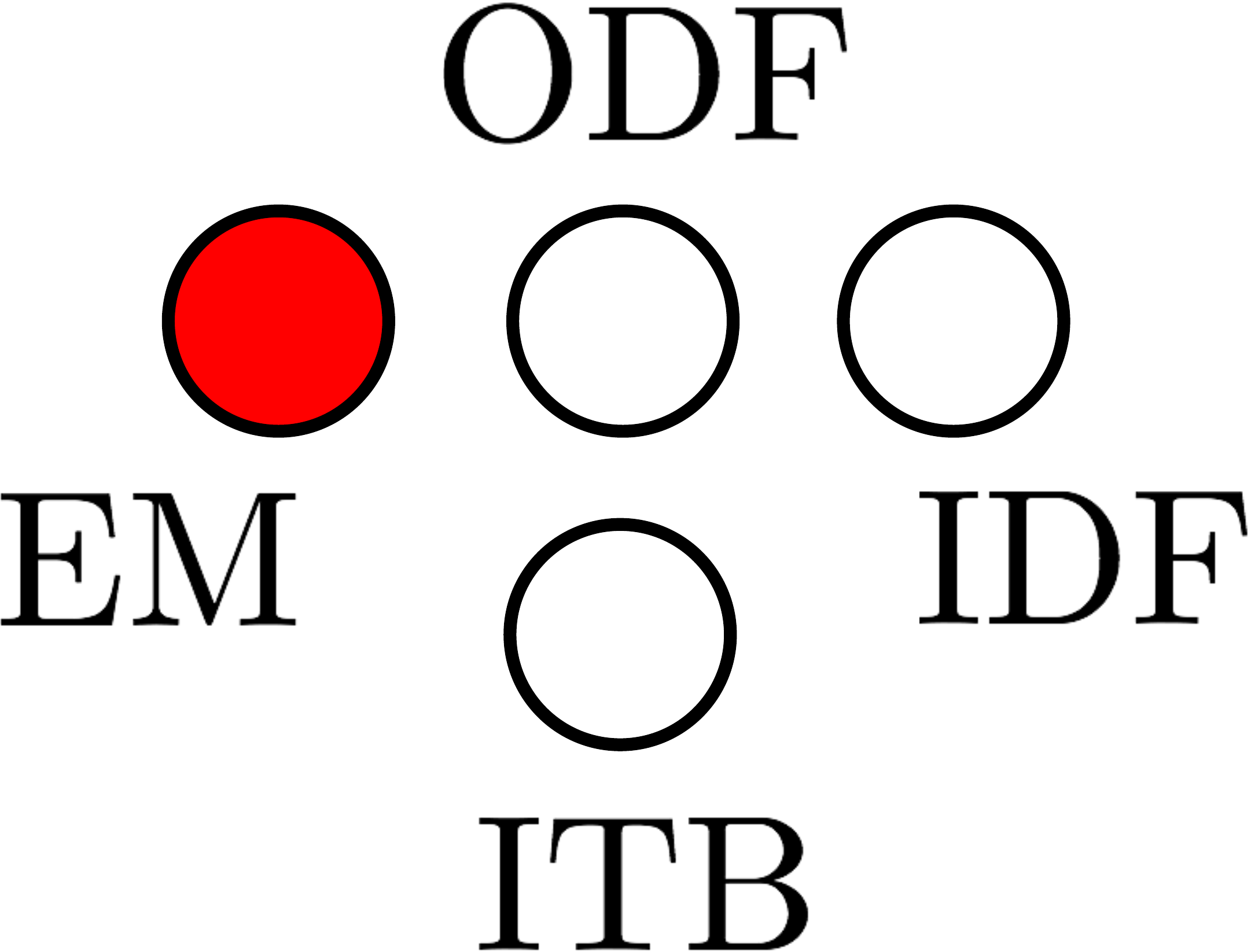}
   &
   \includegraphics[width=0.8\linewidth]{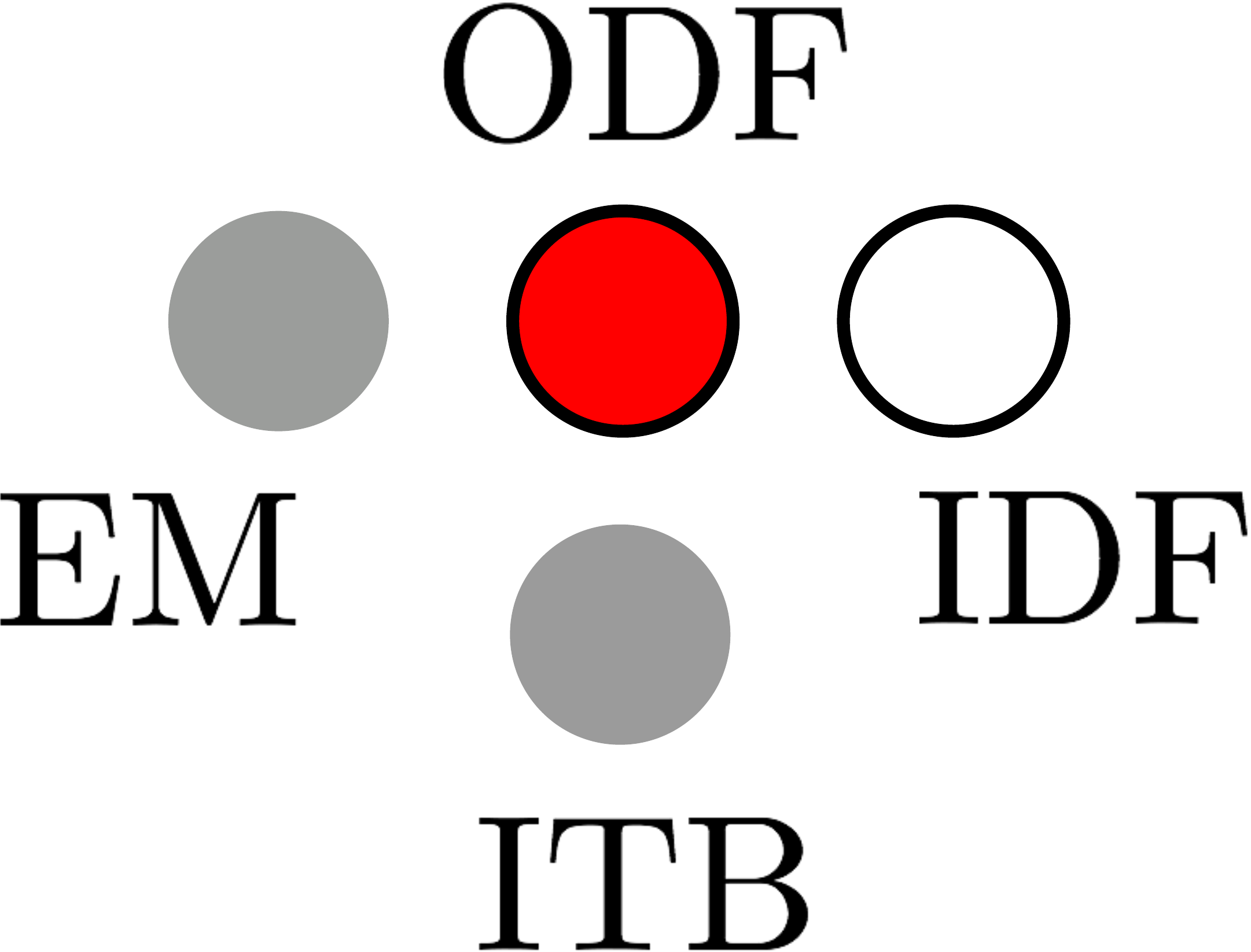}
   & \vspace{0.35cm} 
   \includegraphics[width=0.8\linewidth]{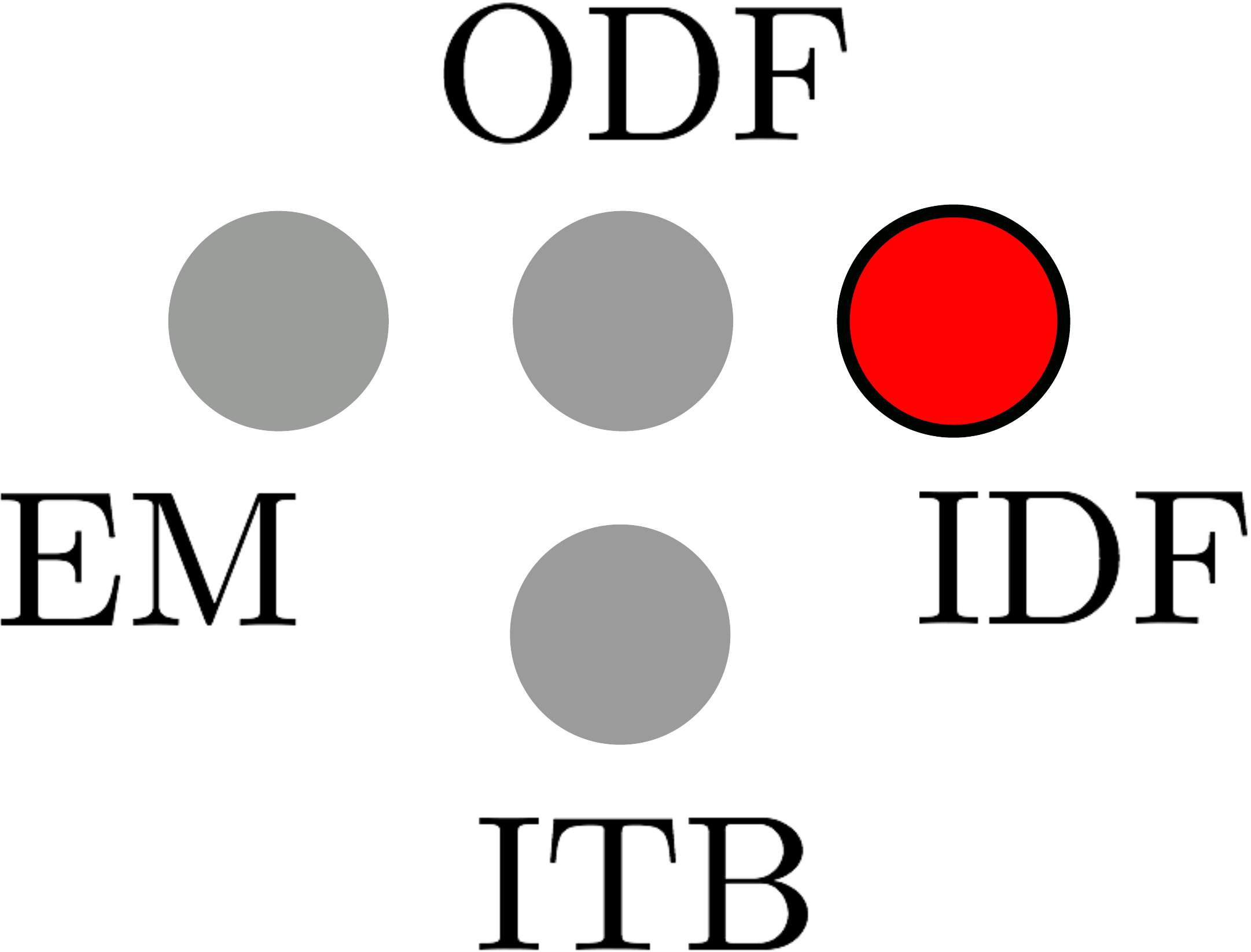}
   \tabularnewline [2.5ex] 
 \hline\hline
 Retarded propagator & $\DissQBM$ (Eq.    \ref{eqDissQBM})& $\PropPh^0$ (\eqnref{eqPropPh})& $\PropNP$ (\eqnref{eqPropNP})  & $\KerEMR$ (\eqnref{GRRaGMA}) & $\PropPo$ (\eqnref{eqPropPo}) & \vspace{0.2cm} $\PropPh$ (\eqnref{PhDFGreen})
  \tabularnewline  [1ex] 
 \hline
 Hadamard propagator & $\NoiQBM$ (\eqnref{NoiseQBMDef}) & $\HadPh^0$ (\eqnref{eqHadPh}) &  $\mathrlap{\diagdown}\diagup$ & $\KerEMH $ (\eqnref{FDRMA}) &  $\mathrlap{\diagdown}\diagup$ & \vspace{0.2cm}$\mathrlap{\diagdown}\diagup$ \tabularnewline  [1ex] 
 \hline
 Compound Noise kernel & $\mathrlap{\diagdown}\diagup$ & $\mathrlap{\diagdown}\diagup$ & $\NoiseNP$ (\eqnref{EqNoiseNP}) & $\mathrlap{\diagdown}\diagup$ & $\KerNoise_{\rm Env}$ (\eqnref{EnvSumKernel}) &  \vspace{0.2cm}$\CorrPh_{\rm Tot}$ (\eqnref{CorrZ}) \tabularnewline  [1ex] 
 \hline
 \end{tabular}
 \caption{Summary of the notation for the different kernels in our calculation. The red coloured circles represent the degree of freedom being traced out, whereas the gray circles represent additional traces taken in a previous step.  The first row shows the retarded propagators, i.e., the dissipation kernels, whereas the noise kernels are split between those which can be expressed as Hadamard propagators (second row) and the more involved compound noise kernels (third row). In the last column, the different notation for $\CorrPh_{\rm Tot}$ arises from the fact that this term represents a fluctuation but not technically a noise source.}\label{Tablepropagators}
\end{table*}

In this appendix we detail the functional integrations over all the internal degrees of freedom in our model (ITB, IDF, and ODF) required to obtain the effective influence action for the EM field and, consequently, its effective equation of motion given in Eq.(\ref{TransformedEffFieldEq}). The structure of our model, schematically depicted in \figref{figModel}, requires a sequential tracing over such degrees of freedom: first, we trace the ITB and the IDF, and only after we use the modified influence action of the ODF to trace it over. This procedure is schematically illustrated in the first three columns of Table \ref{Tablepropagators}.

We start by the trace of the ITB acting over the ODF. Because the ITB-ODF interaction is linear, such trace leads to an effective action for the ODF analogous to that of a brownian particle. Following Ref.  \cite{BreuerPet}, we replace the discrete ITB with a continuous bath characterized by the spectral density $\SpQBM(\omega)$ (\eqnref{spectraldensity}), 
obtaining the following influence action
\begin{equation}
    \begin{split}
        S_{\rm QBM}&[\PoDF,\PoDF']=-\int_{t_{\rm in}}^{t_{\rm f}}d\lambda d\lambda'\PoDF^{-}(\lambda)\cdot
        \\
        &
        \times\!\!\Big[\DissQBM(\lambda-\lambda')\PoDF^{+}(\lambda')\!-\!\frac{i}{2}\NoiQBM(\lambda-\lambda')\PoDF^{-}(\lambda')\!\Big],
    \end{split}\label{InfAcQBM}
\end{equation}
where $\PoDF^{-}\equiv\PoDF'-\PoDF$ and $\PoDF^{+}\equiv(\PoDF+\PoDF')/2$. Note that the above equation has the same form of Eq.(\ref{AccionInfluenciaNucleosDyN}), in this case with dissipation and noise kernels corresponding to the quantum brownian motion (QBM):
\begin{equation}\label{eqDissQBM}
\DissQBM(\lambda-\lambda')=2\int_{0}^{+\infty}d\omega~\SpQBM(\omega)~\sin\left(\omega[\lambda-\lambda']\right),
\end{equation}
\begin{equation}
    \begin{split}
        \NoiQBM(\lambda-\lambda')&=2\int_{0}^{+\infty}d\omega\SpQBM(\omega)
        \\
        &
        \times\coth\left(\frac{\beta_{\CDamp}}{2}\omega\right)\cos\left[\omega(\lambda-\lambda')\right].
    \end{split}\label{NoiseQBMDef}
\end{equation}
This tracing process corresponds to the first column of Table \ref{Tablepropagators}.

The second step is to trace out the IDF. Here, the form of the involved free and interaction actions guarantees that the integrals are Gaussian and thus analytically solvable. After performing them, we obtain a second contribution to the effective influence functional for the ODF, this time due to the IDF:
\begin{equation}\label{influencePol1st}
    \begin{split}
        S_{\rm IF}&[\PoDF,\PoDF']=\int_{t_{\rm in}}^{t_{\rm f}}d\lambda d\lambda'4\MPo\FPo\FPh\CPoPh^{2}\PoDF^{-}(\lambda)\cdot
        \\
        &
        \times\Big[\PropPh^{0}(\lambda-\lambda')\PoDF^{+}(\lambda')+\frac{i}{2}\HadPh
^{0}(\lambda-\lambda')\PoDF^{-}(\lambda')\Big],
    \end{split}
\end{equation}
where $\PropPh
^{0}$ and $\HadPh^{0}$ are the retarded and Hadamard propagators of the free IDF. Such propagators are more easily defined through their Laplace transforms (expressed as functions of the variable $s$ from now on):
\begin{equation}\label{eqPropPh}
\PropPh
^{0}(s)=\frac{1}{\left[s^{2}+\FPh^{2}\right]},
\end{equation}
\begin{equation}\label{eqHadPh}
\HadPh
^{0}(s)=\frac{1}{\FPh}\coth\left(\frac{\beta_{\theta}}{2}\FPh\right)\frac{s}{(s^{2}+\FPh^2)}.
\end{equation}
It is worth noting that the influence action \eqnref{influencePol1st} is analogous to Eq. (\ref{InfAcQBM}) but for an environment composed of a single oscillator (the IDF in this case).

The final step corresponds to the third column of Table \ref{Tablepropagators}, i.e. tracing out the ODF under the extra influence action $S_{\rm QBM}[\PoDF,\PoDF']+S_{\rm IF}[\PoDF,\PoDF']$ (\eqnref{InfAcQBM} and \ref{influencePol1st}). Again, all the actions involved in this step are quadratic and, therefore, the functional integration can be performed exactly. After carrying them out, we obtain the following influence action for the EM field:
\begin{equation}\label{IFEMfield}
    \begin{split}
        S_{\rm IF}\big[A^{\mu}&,A'^{\mu}\big]=
        \\
        &=\int_{t_{\rm in}}^{t_{\rm f}}\!\!d\lambda d\lambda'\frac{\CAPo^{2}}{\MPo}\mathbf{E}^{-}(\lambda,\mathbf{z}_{0})\cdot\!\Big[\PropNP(\lambda-\lambda')
        \\
        &
        \times\mathbf{E}^{+}(\lambda',\mathbf{z}_{0})+\frac{i}{2}\NoiseNP(\lambda,\lambda')\mathbf{E}^{-}(\lambda',\mathbf{z}_{0})\Big].
    \end{split}
\end{equation}
Here, we define the retarded propagator of the NP as that of the ODF obtained after tracing out the IDF and the ITB. This function is defined through its Laplace transform as
\begin{equation}\label{eqPropNP}
\PropNP^{-1}
(s)=\left[s^{2}+\FPo^{2}-\frac{\DissQBMNS_{\gamma}(s)}{\MPo}-2\FPo\FPh\CPoPh^{2}~
\PropPh
^{0}(s)\right].
\end{equation}
In \eqnref{IFEMfield}, the noise kernel of the nanoparticle, $\NoiseNP$, it does not correspond to a Hadamard propagator of the ODF but to a more complex structure involving both the Hadamard propagators of the ITB and the IDF and their retarded propagators:
\begin{widetext}
\begin{equation}
    \begin{split}
        \NoiseNP(\lambda,\lambda')&=\frac{1}{2\FPo}\coth\left[\frac{\beta_{\FPo}}{2}\FPo\right]
        \left[
        \dot{G}_{\rm NP}(\lambda -t_{\rm in})\dot{G}_{\rm NP}(\lambda' -t_{\rm in})\right.
        \hspace{0cm}
        \left.+\FPo^{2}\PropNP(\lambda -t_{\rm in})\PropNP(\lambda' -t_{\rm in})\right]+
        \\
        &
        +
        \int_{t_{\rm in}}^{t_{\rm f}}d\lambda''d\lambda'''\PropNP(\lambda -\lambda'')\PropNP(\lambda'-\lambda''')
        \times
        \Big[2\FPo\FPh\CPoPh^{2}\HadPh^{0}(\lambda''-\lambda''') +
        \frac{1}{2}\NoiQBM(\lambda-\lambda')\Big].
    \end{split}
    \label{EqNoiseNP}
\end{equation}
\end{widetext}

Once the influence functional for the EM field under the influence of the remaining subsystems (\eqnref{IFEMfield}) has been calculated, our goal is to obtain the effective equation of motion of the EM field. In order to do this, it is useful to rewrite \eqnref{IFEMfield} as
\begin{equation}
    \begin{split}
        S_{\rm IF}&\left[A^{\mu},A'^{\mu}\right]=
        \int d\mathbf{x}\int_{t_{\rm in}}^{t_{\rm f}}d\lambda~\delta(\mathbf{x}-\mathbf{z}_{0})\mathbf{E}^{-}(\lambda,\mathbf{x})\cdot
        \\
        &
        \times\int_{t_{\rm in}}^{t_{\rm f}}d\lambda'
        \frac{\CAPo^{2}}{\MPo}\Big[\PropNP(\lambda-\lambda')\mathbf{E}^{+}(\lambda',\mathbf{x}) +
        \\
        &
        \hspace{2.5cm}+
        \frac{i}{2}\NoiseNP(\lambda,\lambda')\mathbf{E}^{-}(\lambda',\mathbf{x})\Big].
    \end{split}
\end{equation}
We now follow Ref. \cite{CTPGauge} by introducing a gauge fixing action for the EM field, where we specifically choose the temporal gauge $A^{0}\equiv 0$ (such that $\mathbf{E}=-\dot{\mathbf{A}}$). In this way, the four equations of motion for each component of $A^{\mu}$ are reduced to three equations for the components of the vector potential, $\mathbf{A}$, plus one residual condition provided by the fourth equation in combination with the chosen gauge. Such a residual condition is referred to as the generalized Coulomb gauge condition  \cite{CTPGauge}. The computation of the effective equation of motion for the EM field, which in this gauge is particularly simple, is now carried out by extreming the CTP action $S_{\rm EM}[A^{\mu}]-S_{\rm EM}[A'^{\mu}]+S_{\rm IF}\left[A^{\mu},A'^{\mu}\right]$ according to  \eqnref{EcMovCTP}. Noting that the condition $\mathbf{A}=\mathbf{A}'$ implies $\mathbf{A}^{-}\equiv 0$ and $\mathbf{A}^{+}=\mathbf{A}$, we can obtain
\begin{equation}
    \begin{split}
        \nabla&\times(\nabla\times \mathbf{A})+\frac{1}{c^{2}}\frac{\partial^{2}\mathbf{A}}{dt^{2}}+\delta(\mathbf{x}-\mathbf{z}_{0})
        \\
        &
        \times\!\frac{d}{dt}\!\left[\int_{t_{\rm in}}^{t}\!d\lambda\frac{\CAPo^{2}}{\MPo c^{2}\epsilon_{0}}\PropNP
(t-\lambda)\dot{\mathbf{A}}(\lambda,\mathbf{x})\right]=0,
    \end{split}
\end{equation}
or, after transforming to Laplace space,
\begin{equation}
\begin{split}
\nabla&\times(\nabla\times\mathbf{A})+
\\
&+\frac{s^{2}}{c^{2}}\left[1+\delta(\mathbf{x}-\mathbf{z}_{0})\frac{\CAPo^{2}
\PropNP
(s)}{\MPo\epsilon_{0}}\right]\mathbf{A}(s,\mathbf{x})=\mathbf{F}_{0}.
\end{split}\label{curlAprov}
\end{equation}
Here, the r.h.s. of the equation contains all the terms dependent on the initial conditions of the field, which we do not write explicitly since they are unnecessary for determining the polarizability of our model.
Noting that the propagators in \eqnref{curlAprov} are causal, we can easily transform to Fourier space by substituting $s=-i\omega$ and, after setting $\mathbf{z}_{0}\equiv 0$, directly obtain \eqnref{TransformedEffFieldEq} of the main text.

\section{Tracing out the EM field}\label{TEMF}

The remaining appendices are devoted to the calculation of the internal energy $u(t)$ in the main text, by means of calculating the generating functional (this functional is given in \eqnref{GeneratingFunctionalPathIntegral}). The first step in this procedure is to trace both the EM and the ITB environments (first and fourth columns in Table \ref{Tablepropagators}). Note that the latter trace has already been undertaken in Appendix \ref{TOIDF}. This appendix is thus devoted to taking the trace over the EM field, a procedure involving some delicate steps.

To trace out the EM field, we first separate all the factors and integrations involving the EM field on Eq.(\ref{GeneratingFunctionalPathIntegral}), the resulting object being the influence functional of the EM field over the rest of the system, given by:
\begin{equation}
    \begin{split}
        \mathcal{F}_{\rm EM}&\left[J_{\mu},J'_{\mu}\right]\!=\!\!\int 
        \!\!dA^{\mu}_{\rm f}\!\int \!\!dA^{\mu}_{\rm in}dA'^{\mu}_{\rm in}\rho_{\rm EM}(A^{\mu}_{\rm in},A'^{\mu}_{\rm in};t_{\rm in})
        \\
        &
        \times\!\!\int^{A^{\mu}_{\rm f}}_{A^{\mu}_{\rm in}}\!\mathcal{D}A^{\mu}\int^{A^{\mu}_{\rm f}}_{A'^{\mu}_{\rm in}}\!\mathcal{D}A'^{\mu} e^{i\left(S_{\rm EM}\left[A^{\mu}\right]+S_{\rm Dip}\left[J_{\mu},A^{\mu}\right]\right)}
        \\
        &
        \times
        e^{-i\left(S_{\rm EM}\left[A'^{\mu}\right]+S_{\rm Dip}\left[J'_{\mu},A'^{\mu}\right]\right)}.
    \end{split}
\end{equation}
In order to perform the path integrations, it is important to note that, due to the gauge symmetry of the EM theory, such integrations must be taken over one class of paths for the EM field, i.e., redundant paths connected through gauge transformations must be excluded from the integration. Formally, this is done by introducing the so-called Faddeev-Popov procedure in the calculations  \cite{CTPGauge}. Nevertheless, regardless on the chosen gauge, the CTP-integral is Gaussian, so we can perform it analytically to obtain:
\begin{equation}
    \begin{split}
        \mathcal{F}&_{\rm EM}\left[J^{\mu-},J^{\mu+}\right] = \exp\bigg\lbrace i \int d^{4}y\int d^{4}y'J^{\mu-}(y)
        \\
        &
        \times
        \Big[\PropEM^{\rm Ret}(y,y')J^{\nu+}(y')+\frac{i}{4}\PropEM^{\rm H}(y,y')J^{\nu-}(y')\Big]\bigg\rbrace,
    \end{split}\label{InfluenceFunctionalEM}
\end{equation}
where $J^{\nu+}=(J^{\nu}+J'^{\nu})/2$ and $J^{\nu-}=J^{\nu}-J'^{\nu}$. The tensors  $\PropEM^{\rm Ret}$ and $\PropEM^{\rm H}$ stand respectively for the retarded and Hadamard propagators of the EM field in the chosen state, which we assume is a thermal state with temperature $T_{\rm EM}$. Note that, whereas \eqnref{InfluenceFunctionalEM} is valid for every gauge, the expression of the propagators is not. In our case, it is convenient to fix the Feynman gauge, where $\PropEM^{\rm Ret}$ and $\PropEM^{\rm H}$ can be expressed in terms of the massless scalar field propagators $\PropScalar_{\rm Ret}$ and $\PropScalar_{\rm H}$  \cite{BehuninHu2009},
\begin{equation}
\PropEM^{\rm Ret,H}(y,y')=\frac{1}{\epsilon_{0}}\eta_{\mu\nu}\PropScalar_{\rm Ret,H}(y,y').
\label{PropagatorsEMFeynmanGauge}
\end{equation}
Such scalar field propagators are given by
\begin{equation}\label{RetKernelEM}
    \begin{split}
        \PropScalar&_{\rm Ret}(y,y')\equiv\PropScalar_{\rm Ret}(y-y') =
        \\
        &
        =\int\frac{d^{4}p}{(2\pi)^{4}}\frac{e^{-ip(y-y')}}{\left(p_{0}+i\epsilon\right)^{2}-c^{2}\mathbf{p}^{2}}
        =
        \\
        &
        =
        -\int\frac{d\mathbf{p}}{(2\pi)^{3}}e^{i\mathbf{p}\cdot\left(\mathbf{y}-\mathbf{y}'\right)}\theta(t-t')\frac{\sin\left[\omega_{\mathbf{p}}(t-t')\right]}{\omega_{\mathbf{p}}},
    \end{split}
\end{equation}
and
\begin{equation}\label{HKernelEM}
    \begin{split}
        &\PropScalar_{\rm H}(y,y')\equiv\PropScalar_{\rm H}(y-y') = 
        \\
        &
        = \!\!\int\!\!\frac{d^{4}p}{(2\pi)^{3}}e^{-ip(y-y')}\delta(p_{0}^{2}-c^{2}\mathbf{p}^{2})\!\coth\!\!\left[\frac{\beta_{\rm EM}}{2}|p_{0}|\right]
        \\
        &
        =\!\!
        \int\!\!\frac{d\mathbf{p}}{(2\pi)^{3}}e^{i\mathbf{p}\cdot\left(\mathbf{y}-\mathbf{y}'\right)}\!\coth\!\!\left[\frac{\beta_{\rm EM}}{2}\omega_{\mathbf{p}}\right]\!\frac{\cos\left[\omega_{\mathbf{p}}(t-t')\right]}{\omega_{\mathbf{p}}},
    \end{split}
\end{equation}
where $\epsilon$ stands for the small parameter of Feynman prescription for propagators, $y_{0}=ct$, and $\omega_{\mathbf{p}}=c|\mathbf{p}|$.

It is important to note that $\mathcal{F}_{\rm EM}$ is a functional of $\PoDF$, $\PoDF'$ and also a function of $\mathbf{z}_{0}$ through its dependence on the four-currents $J^{\mu}$. We can express this combined dependence as $\mathcal{F}_{\rm EM}\left[\PoDF^{-},\PoDF^{+},\right.\left.\mathbf{z}_{0}\right)$. Note that, in the forthcoming section, we will be interested in tracing out the degrees of freedom of the ODF, $\PoDF$ and $\PoDF'$. Thus, it is convenient to write the influence functional \eqnref{InfluenceFunctionalEM} in a form where the dependence with these variables is explicit. This is easily achieved by using the definition of the currents (\eqnref{defcurrents}) to write the influence functional as
\begin{widetext}
\begin{equation}\label{FEMprov}
\mathcal{F}_{\rm EM}\left[\PoDF^{-},\PoDF^{+},\right.\left.\mathbf{z}_{0}\right)=\exp\bigg\{i
\int_{t_{\rm in}}^{t_{\rm f}} d\lambda d\lambda'\PoDFc^{j-}(\lambda)\Big[2\MKerEMR(\lambda,\lambda';\mathbf{z}_{0})\PoDFc^{k+}(\lambda')+\frac{i}{2}\MKerEMH(\lambda,\lambda';\mathbf{z}_{0})\PoDFc^{k-}(\lambda')\Big]\bigg\}.
\end{equation}
The above recasting of the influence functional defines yet another set of retarded and noise propagators,
\begin{equation}\label{propagatornew1}
\MKerEMR(\lambda,\lambda';\mathbf{z}_{0})
\equiv
\MKerEMR(\lambda-\lambda')
=\frac{\CAPo^{2}}{2}\left[\left(-\delta_{j}^{\mu}\partial_{0}+c\delta_{0}^{\mu}\partial_{j}\right)\left(-\delta_{k}^{\nu}\partial'_{0}+c\delta_{0}^{\nu}\partial'_{k}\right)\PropEM^{\rm Ret}(y,y')\right]\Big|_{y^{\alpha}=z_{0}^{\alpha},y'^{\alpha}=z'^{\alpha}_{0}},
\end{equation}
\begin{equation}\label{propagatornew2}
\MKerEMH(\lambda,\lambda';\mathbf{z}_{0})
\equiv
\MKerEMH(\lambda-\lambda')
=\frac{\CAPo^{2}}{2}\left[\left(-\delta_{j}^{\mu}\partial_{0}+c\delta_{0}^{\mu}\partial_{j}\right)\left(-\delta_{k}^{\nu}\partial'_{0}+c\delta_{0}^{\nu}\partial'_{k}\right)\PropEM^{\rm H}(y,y')\right]\Big|_{y^{\alpha}=z_{0}^{\alpha},y'^{\alpha}=z'^{\alpha}_{0}}.
\end{equation}
Here, the simplification from the variables $\lambda,\lambda'$ to the difference $\lambda-\lambda'$ originates from the property $\PropEM^{\rm Ret,H}(y,y')\equiv\PropEM^{\rm Ret,H}(y-y')$ of the original propagators.

The propagators defined by \eqnref{propagatornew1} and \ref{propagatornew2} can be analytically calculated. For the kernel associated to the retarded propagator, we have
\begin{equation}
    \begin{split}
        \MKerEMR(\lambda-\lambda')&=
        \frac{\CAPo^{2}}{2\epsilon_{0}}\left[\int\frac{d\mathbf{p}}{(2\pi)^{3}}\left[\partial_{0}\partial'_{0}\delta_{jk}-c^{2}p_{j}p_{k}\right]\theta(t-t')\frac{\sin\left[\omega_{\mathbf{p}}(t-t')\right]}{\omega_{\mathbf{p}}}\right]\Big|_{t=\lambda,t'=\lambda'}
        =
        \\
        &
        =-\frac{\CAPo^{2}}{2\epsilon_{0}}\delta_{jk}\delta(\lambda-\lambda')\int\frac{d\mathbf{p}}{(2\pi)^{3}}+\frac{\CAPo^{2}}{2\epsilon_{0}}\delta_{jk}\theta(\lambda-\lambda')\int\frac{d\mathbf{p}}{(2\pi)^{3}}\left[\omega_{\mathbf{p}}^{2}-c^{2}p^{2}_{j}\right]~\frac{\sin\left[\omega_{\mathbf{p}}(\lambda-\lambda')\right]}{\omega_{\mathbf{p}}}
        \\
        &
        =-\frac{\CAPo^{2}}{2c^{3}\epsilon_{0}}\delta_{jk}\delta(\lambda-\lambda')\int_{0}^{+\infty}\frac{d\omega_{\mathbf{p}}}{2\pi}\frac{\omega_{\mathbf{p}}^{2}}{\pi}
        +\frac{\CAPo^{2}}{3c^{3}\epsilon_{0}}\delta_{jk}\theta(\lambda-\lambda')\int_{0}^{+\infty}\frac{d\omega_{\mathbf{p}}}{2\pi}\frac{\omega_{\mathbf{p}}^{3}}{\pi}\sin\left[\omega_{\mathbf{p}}(\lambda-\lambda')\right],
    \end{split}\label{GjkEM}
\end{equation}
where in the second line we have used the fact that the off-diagonal terms are zero, and in the third line we have computed the angular integrations in spherical coordinates. 
Note that the first term in the last line of \eqnref{GjkEM} is divergent and should be included in a frequency renormalization of the ODF as detailed below. On the other hand, the second term represents the causality kernel that generates dissipation on the ODF, and is given also by a divergent integral. Such a divergence originates from the consideration of the body as a static point (i.e., with its center of mass being not in motion), the resulting interaction with the EM field occurring at a single spatial point.
 In practice, the divergent behavior is prevented by introducing a frequency cut-off function $\EMfCO(\omega_{\mathbf{p}})$ fulfilling $\EMfCO(\omega_{\mathbf{p}})\rightarrow 0$ for $\omega_{\mathbf{p}}\gtrapprox\EMCO$, which accounts for the fact that the body does not interact with EM modes above the cut-off frequency $\EMCO$. We can thus finally write the dissipation kernel as
\begin{equation}
\MKerEMR(\lambda-\lambda')=-\frac{\CAPo^{2}\delta_{jk}}{2c^{3}\epsilon_{0}}\delta(\lambda-\lambda')\int_{0}^{+\infty}\frac{d\omega_{\mathbf{p}}}{2\pi}\omega_{\mathbf{p}}^{2}\frac{\EMfCO(\omega_{\mathbf{p}})}{\pi}+\frac{\CAPo^{2}\delta_{jk}}{3c^{3}\epsilon_{0}}\theta(\lambda-\lambda')\int_{0}^{+\infty}\frac{d\omega_{\mathbf{p}}}{2\pi}\omega_{\mathbf{p}}^{3}\frac{\EMfCO(\omega_{\mathbf{p}})}{\pi}\sin\left[\omega_{\mathbf{p}}(\lambda-\lambda')\right].
\label{DRet}
\end{equation}
In part of the literature, the cutoff function stems from the interpretation of the quantity $\CAPo^{2}\EMfCO(\omega_{\mathbf{p}})$ as the spectral density associated to the EM field, since it characterizes its interaction properties as an environment.

Let us now focus on the kernel associated to the Hadamard propagator of the EM field. Introducing the same cutoff function as above and assuming it is an even function of $\omega_{\mathbf{p}}$, we can write such kernel as
\begin{equation}
    \begin{split}
        \MKerEMH(\lambda-\lambda')&=\frac{\CAPo^{2}}{3c^{3}\epsilon_{0}}~\delta_{jk}\int_{0}^{+\infty}\frac{d\omega_{\mathbf{p}}}{2\pi}~\omega_{\mathbf{p}}^{3}~\frac{\EMfCO(\omega_{\mathbf{p}})}{\pi}~\coth\left(\frac{\beta_{\rm EM}}{2}\omega_{\mathbf{p}}\right)\cos\left[\omega_{\mathbf{p}}(\lambda-\lambda')\right] 
        \\
        &
        = \frac{\CAPo^{2}}{6c^{3}\epsilon_{0}}~\delta_{jk}\int_{-\infty}^{+\infty}\frac{d\omega_{\mathbf{p}}}{2\pi}~\omega_{\mathbf{p}}^{3}~\frac{\EMfCO(\omega_{\mathbf{p}})}{\pi}~\coth\left(\frac{\beta_{\rm EM}}{2}\omega_{\mathbf{p}}\right)~e^{-i\omega_{\mathbf{p}}(\lambda-\lambda')}
    \end{split}\label{DH}
\end{equation}
Importantly, both kernels $\MKerEMR$ and $\MKerEMH$ are diagonal and proportional to the identity operator, i.e. $\MKerEMR\equiv \delta_{jk}\KerEMR$, $\MKerEMH\equiv\delta_{jk}\KerEMH$. This allows us to simplify the influence functional \eqnref{FEMprov} to
\begin{equation}
\mathcal{F}_{\rm EM}\left[\PoDF^{-},\PoDF^{+}\right]=\exp\Big\{i
\int_{t_{\rm in}}^{t_{\rm f}} d\lambda d\lambda'~\PoDF^{-}(\lambda)\cdot\Big[2\KerEMR(\lambda-\lambda')~\PoDF^{+}(\lambda')+\frac{i}{2}\KerEMH(\lambda-\lambda')~\PoDF^{-}(\lambda')\Big]\Big\}.
\label{FEMZ0}
\end{equation}
\end{widetext}
It is clear that each cartesian component of the ODF degrees of freedom $\PoDF^{-}(\lambda)$ is uncoupled from the others, a fact that will simplify the following calculations.

\subsection*{Radiation reaction and (weak coupling) Markov approximation}\label{AppRR}

So far, we have traced out the EM field, obtaining the influence functional, \eqnref{FEMZ0}, describing its effect on the ODF. Our next step towards the calculation of the internal energy should be to trace out the ODF including both such influence functional and the one corresponding to the ITB. However, as we will see below, the effect of the EM field on the ODF dynamics presents mathematical problems that have to be taken into account first. In this Appendix we detail such problems and the weak-coupling approximation that we use to solve them.


Let us derive the equation of motion for the ODF under the influence of the EM field. Since, according to \eqnref{FEMZ0}, each cartesian component of $\mathbf{x}_\Omega$ is uncoupled from the rest, we can treat them separately. The CTP action for component $j$ can thus be written as
\begin{equation}
\begin{split}
    S_{\rm eff}&[\PoDFc^{j},\PoDFc'^{j}]=S_{\FPo}[\PoDFc^{j}]-S_{\FPo}[\PoDFc'^{j}]+
    \\
    &
    +\int_{t_{\rm in}}^{t_{\rm f}} d\lambda d\lambda'\PoDFc^{j-}(\lambda)\Big[2\KerEMR(\lambda-\lambda')\PoDFc^{j+}(\lambda')+
    \\
    &
    \hspace{2.5cm}+\frac{i}{2}\KerEMH(\lambda-\lambda')\PoDFc^{j-}(\lambda')\Big].
\end{split}
\end{equation}
From here, the semiclassical equation of motion for the ODF can be obtained by extreming the effective action,
\begin{equation}
    \frac{\delta S_{\rm eff}[\PoDFc^{j},\PoDFc'^{j}]}{\delta \PoDFc^{j-}(t)}\Big|_{\PoDFc^{j-}=0}=0,
\end{equation}
which yields
\begin{equation}
\begin{split}
\ddot{x}_\Omega&^{j}(t)+\FPo^{2}\PoDFc^{j}(t)-
\\
&
-\frac{2}{\MPo}\int_{t_{\rm in}}^{t}d\lambda \KerEMR(t-\lambda)~\PoDFc^{j}(\lambda)=0.
\end{split}\label{EffectiveEqCTP}
\end{equation}
If we now introduce above the expression for the retarded kernel, Eq.(\ref{DRet}), we find that its first term results in a renormalization of the frequency $\FPo$, while its second term acts as the \emph{radiation reaction kernel}, i.e.,
\begin{equation}\label{SemiClassicalEqMotionQ}
    \begin{split}
        \ddot{x}&_\Omega^{j}(t) +
        \\
        &
        +\left[\FPo^{2}+\frac{\CAPo^{2}}{\MPo c^{3}\epsilon_{0}}\int_{0}^{+\infty}\!\!\frac{d\omega_{\mathbf{p}}}{2\pi}\omega_{\mathbf{p}}^{2}\frac{\EMfCO(\omega_{\mathbf{p}})}{\pi}\right]\PoDFc^{j}(t)-
        \\
        &
        -\frac{2}{\MPo}\int_{t_{\rm in}}^{t}d\lambda\KerEMR^{\rm RR}(t-\lambda)\PoDFc^{j}(\lambda) = 0.
    \end{split}
\end{equation}
Let us now rewrite the above equation in a more transparent way. Noting that the radiation reaction kernel, $\KerEMR^{\rm RR}$, is defined as
\begin{widetext}
\begin{equation}
\begin{split}
    &\KerEMR^{\rm RR}(t-\lambda)=
    \\
    &=\frac{\CAPo^{2}}{3c^{3}\epsilon_{0}}\theta(t-\lambda)\int_{0}^{+\infty}\frac{d\omega_{\mathbf{p}}}{2\pi}~\omega_{\mathbf{p}}^{3}\frac{\EMfCO(\omega_{\mathbf{p}})}{\pi}\sin\left[\omega_{\mathbf{p}}(t-\lambda)\right]=\frac{\CAPo^{2}}{3c^{3}\epsilon_{0}}\theta(t-\lambda)\frac{\partial^{3}}{\partial t^{3}}\left[\int_{0}^{+\infty}\frac{d\omega_{\mathbf{p}}}{2\pi}\frac{\EMfCO(\omega_{\mathbf{p}})}{\pi}\cos\left[\omega_{\mathbf{p}}(t-\lambda)\right]\right],
\end{split} \label{GEMRRdef}   
\end{equation}
the convolution in \eqnref{SemiClassicalEqMotionQ} can be recast as
\begin{equation}
-\frac{2}{\MPo}\int_{t_{\rm in}}^{t}d\lambda~\KerEMR^{\rm RR}(t-\lambda)~\PoDFc^{j}(\lambda)=-\frac{2}{\MPo}\frac{d}{dt}\left[\int_{t_{\rm in}}^{t}d\lambda~\DampKRR(t-\lambda)~\PoDFc^{j}(\lambda)\right]-\frac{2\CAPo^{2}}{3\MPo c^{3}\epsilon_{0}}\int_{0}^{+\infty}\frac{d\omega_{\mathbf{p}}}{2\pi}~\omega_{\mathbf{p}}^{2}~\frac{\EMfCO(\omega_{\mathbf{p}})}{\pi}~\PoDFc^{j}(t).
\label{DRetRR}
\end{equation}
\end{widetext}
The second term on the r.h.s. above can be reabsorbed in the renormalization of the frequency $\Omega$. This would result in two renormalization terms in \eqnref{SemiClassicalEqMotionQ}, which can always be erased by including a counter-term in the initial ODF actions with no repercussion in the dynamics. We will thus omit these terms hereafter, so that the equation of motion simplifies to
\begin{equation}
\begin{split}
\ddot{x}_\Omega^{j}(t)&+\FPo^{2}\PoDFc^{j}(t)-
\\
&-\frac{2}{\MPo}\frac{d}{dt}\left[\int_{t_{\rm in}}^{t}d\lambda~\DampKRR(t-\lambda)\PoDFc^{j}(\lambda)\right]=0.
\end{split}
\label{SemiClassicalEqMotionQGamma}
\end{equation}
The above equation contains the \emph{radiation reaction damping kernel} $\DampKRR$, which can be analytically computed as
\begin{equation}
    \begin{split}
        \DampKRR(t-\lambda)&=\frac{\CAPo^{2}}{3c^{3}\epsilon_{0}}\theta(t-\lambda)\frac{\partial^{2}}{\partial t^{2}} \\
        &
        \\
        &\times \int_{0}^{+\infty}\frac{d\omega_{\mathbf{p}}}{2\pi}\frac{\EMfCO(\omega_{\mathbf{p}})}{\pi}\cos\left[\omega_{\mathbf{p}}(t-\lambda)\right]=
        \\
        &
        =\frac{\CAPo^{2}}{6\pi c^{3}\epsilon_{0}}~\theta(t-\lambda)~\delta^{''}(t-\lambda),
    \end{split}\label{GammaRadiationReaction}
\end{equation}
where in the last term we take the cutoff function equal to $1$ for simplicity, since this has no crucial impact on our final argument.

%
%
%
Finally, introducing Eq.(\ref{GammaRadiationReaction}) into the semi-classical equation of motion Eq.(\ref{SemiClassicalEqMotionQGamma}), we simplify it to
\begin{equation}
\ddot{x}_\Omega^{j}(t)+\FPo^{2}~\PoDFc^{j}(t)-\DampCRR\dddot{x}_\Omega^{j}(t)=0,
\label{RadiationReactionEq}
\end{equation}
where the \emph{radiation reaction coefficient} is defined as $\DampCRR=\CAPo^{2}/(6\pi \MPo c^{3}\epsilon_{0})$. 
This is the well-known radiation reaction equation of motion, i.e., the equation describing a dipole interacting with an EM field in for any value of the coupling \cite{Milonni}. The solutions of such equation, however, are known to present several problems due to `runaways' and `preaccelerations' which result in causality violations. These inconsistencies originate from the punctual nature of the considered interaction \cite{Milonni}.

There are two ways in which one can get around the runaway problem. First, including in our problem the non-punctual nature of the dipole. The second is to perform a weak coupling approximation (also labeled Markovian in the literature). Given that our results in the main text show that the coupling between the EM field and the ODF is indeed weak, we will opt here for the latter.
We start by noticing that the radiation reaction kernel in \eqnref{GammaRadiationReaction} is always a sharply peaked function of its argument, since the cutoff function $\EMfCO(\omega_{\mathbf{p}})$ is assumed to be very broad in frequencies. Thus, in Eq. 
(\ref{SemiClassicalEqMotionQGamma}), the integral in $\lambda$ is effectively restricted to a narrow peak around $\lambda = t$. In this small time window, if the coupling between EM field and dipole is weak, we can approximate the full time evolution of $\PoDF$ by its free evolution,
\begin{widetext}
\begin{equation}
\PoDFc^{j}(\lambda)\approx \PoDFc^{j}(t_{\rm in})\cos\left[\FPo(\lambda-t_{\rm in})\right]+
+\dot{x}_\Omega^{j}(t_{\rm in})\frac{\sin\left[\FPo(\lambda-t_{\rm in})\right]}{\FPo}=
\cos\left[\FPo(\lambda-t)\right]\PoDFc^{j}(t)+
\frac{\sin\left[\FPo(\lambda-t)\right]}{\FPo}\dot{x}_\Omega^{j}(t).
\end{equation}
\end{widetext}
The above free evolution is simply the one of a harmonic oscillator of natural frequency $\FPo$, first expressed in terms of its initial conditions and later in terms of its final conditions.
By introducing the latter form into the convolution term of \eqnref{SemiClassicalEqMotionQGamma}, together with the expression for the radiation reaction damping kernel \eqnref{GammaRadiationReaction}, we obtain
\begin{equation}
\begin{split}
-\frac{2}{\MPo}\frac{d}{dt}\bigg[\int_{t_{\rm in}}^{t}d\lambda \DampKRR(t-\lambda)&\PoDFc^{j}(\lambda)\bigg]\approx
\\
&
\approx\frac{\CAPo^{2}\FPo^{2}}{6\pi \MPo c^{3}\epsilon_{0}}\dot{x}_\Omega^{j}(t).
\end{split}
\end{equation}
With this result, the final equation of motion in the (weak coupling) Markov approximation yields
\begin{equation}
\ddot{x}_\Omega^{j}(t)+\FPo^{2}~\PoDFc^{j}(t)+\frac{\CAPo^{2}\FPo^{2}}{6\pi \MPo c^{3}\epsilon_{0}}~\dot{x}_\Omega^{j}(t)=0,
\label{MarkovApproxEq}
\end{equation}
which is equivalent to the equation of motion of a damped harmonic oscillator of frequency $\FPo^{2}$ and damping constant $\CAPo^{2}\FPo^{2}/(6\pi \MPo c^{3}\epsilon_{0})$. Equation (\ref{MarkovApproxEq}) presents no problems with either causality, preaccelerations, or runaways. 

Although we have undertaken the weak coupling approximation in the equation of motion, it is convenient for our calculations to understand it in terms of the EM field propagators. It is well known that such approximation acts effectively as a change in the nature of the EM bath, from a supraohmic to an ohmic environment. 
This can be observed in the effective change suffered by the retarded kernel under the weak coupling approximation. To see such change, we first note that we can define an effective retarded kernel under the Markov approximation, $\KerEMR^{\rm MA}(t-\lambda)$, by writing the dissipation term in Eq.(\ref{MarkovApproxEq}) as a convolution: 
\begin{equation}
\frac{\CAPo^{2}\FPo^{2}}{6\pi\MPo c^{3}\epsilon_{0}}\dot{x}_\Omega^{j}(t)=-\frac{2}{\MPo}\int_{t_{\rm in}}^{t}\!\!d\lambda\KerEMR^{\rm MA}(t-\lambda)\PoDFc^{j}(\lambda),
\end{equation}
with 
\begin{equation}
\KerEMR^{\rm MA}(t-\lambda)\equiv-\frac{\CAPo^{2}\FPo^{2}}{6\pi c^{3}\epsilon_{0}}~\delta'(t-\lambda).
\end{equation} 
In order to compare with the full propagator (i.e. without the Markov approximation), it is convenient to write the above equation in Fourier space as 
\begin{equation}
\FouKerEMR^{\rm MA}(\omega_{\mathbf{p}})=i\omega_{\mathbf{p}}~\frac{\CAPo^{2}\FPo^{2}}{6\pi c^{3}\epsilon_{0}}=i~{\rm Im}\left[\FouKerEMR^{\rm MA}(\omega_{\mathbf{p}})\right].
\end{equation}
Note that we denote the Fourier transform by a bar over the transformed function, i.e.,  $\overline{G}(\omega)$.
On the other hand, from the definition of the full propagator $\KerEMR^{\rm RR}$ on Eq.(\ref{GEMRRdef}), we can directly express its Fourier transform as
\begin{equation}
\FouKerEMR^{\rm RR}(\omega_{\mathbf{p}})=i\omega_{\mathbf{p}}^{3}~\frac{\CAPo^{2}\EMfCO(\omega_{\mathbf{p}})}{6\pi c^3\epsilon_0}=i~{\rm Im}\left[\FouKerEMR^{\rm RR}(\omega_{\mathbf{p}})\right],
\label{DRetRRFourier}
\end{equation}
where in the last step we have assumed that $\EMfCO(\omega_{\mathbf{p}})$ is real. From the comparison of the above two equations it becomes indeed evident that the effect of the weak coupling approximation is to change from a super-Ohmic to an Ohmic spectral density in Fourier space, i.e., 
\begin{equation}
\begin{split}
\FouKerEMR^{\rm RR}(\omega_{\mathbf{p}})&=i\omega_{\mathbf{p}}^{3}~\frac{\CAPo^{2}}{6\pi c^{3}\epsilon_{0}}\longrightarrow
\\
&
\longrightarrow i \omega_{\mathbf{p}}~\frac{\CAPo^{2}\FPo^{2}}{6\pi c^{3}\epsilon_{0}}=\FouKerEMR^{\rm MA}(\omega_{\mathbf{p}}).
\end{split}
\label{GRRaGMA}
\end{equation}

Importantly, the change in the retarded propagator described above is not the only modification to take into consideration. Indeed, the Hadamard propagator also has to undergo a change under the weak coupling approximation, since both are linked by a fluctuation-dissipation relation (FDR). Specifically, by combining Eqs. (\ref{DH}) and (\ref{DRetRRFourier}), it is straightforward to
prove that such FDR reads
\begin{equation}
\FouKerEMH(\omega_{\mathbf{p}})=\coth\left[\frac{\beta_{\rm EM}\omega_{\mathbf{p}}}{2}\right]{\rm Im}\left[\FouKerEMR^{\rm RR}(\omega_{\mathbf{p}})\right].
\end{equation}
In other words, for a thermal state of the EM field, its Hadamard and retarded kernels are related in Fourier space by the equation above. This is a general property of propagator of the EM field in a thermal state, and thus has to hold also within the weak coupling approximation. Hence, we enforce such relation by performing the substitution of the retarded propagator by its Markovian version in the equation above, i.e.,
\begin{widetext}
\begin{equation}
\FouKerEMH(\omega_{\mathbf{p}})=\coth\left[\frac{\beta_{\rm EM}\omega_{\mathbf{p}}}{2}\right]{\rm Im}\left[\FouKerEMR^{\rm RR}(\omega_{\mathbf{p}})\right]\longrightarrow\coth\left[\frac{\beta_{\rm EM}\omega_{\mathbf{p}}}{2}\right]{\rm Im}\left[\FouKerEMR^{\rm MA}(\omega_{\mathbf{p}})\right]=\FouKerEMH^{\rm MA}(\omega_{\mathbf{p}}).
\label{FDRMA}
\end{equation}
The above equality defines the Hadamard propagator for the EM field under the weak coupling approximation. These two propagators, $\FouKerEMR^{\rm MA}$ and $\FouKerEMH^{\rm MA}$, are the ones we will use to calculate the internal energy in the main text. The corresponding influence action in the Markov approximation is obtained by direct substitution of the EM propagators by their Markovian counterparts in 
Eq.(\ref{FEMZ0}), obtaining
\begin{equation}
\mathcal{F}_{\rm EM}\!\left[\PoDF^{-},\PoDF^{+}\right]\!\longrightarrow\mathcal{F}_{\rm EM}^{\rm MA}\left[\PoDF^{-},\PoDF^{+}\right]\!=\exp\!\left\lbrace \!i
\int_{t_{\rm in}}^{t_{\rm f}} d\lambda d\lambda'\!\PoDF^{-}(\lambda)\!\cdot\!\left[2\KerEMR^{\rm MA}(\lambda-\lambda')\PoDF^{+}(\lambda')+\frac{i}{2}\KerEMH^{\rm MA}(\lambda-\lambda')\PoDF^{-}(\lambda')\right]\right\rbrace,
\label{FEMMA}
\end{equation}
\end{widetext}
after discarding the frequency renormalization terms.

\section{Tracing out the optical degrees of freedom}\label{TOODF}

So far we have undertaken the first two steps toward the computation of the internal energy, i.e. tracing out the ITB in Appendix \ref{TOIDF}, and the EM field in Appendix  \ref{TEMF}. These processes correspond to the first and fourth columns of Table \ref{Tablepropagators}. This Appendix is devoted to the following step, namely tracing out the ODF under the influence of the environments already traced (fifth column in Table \ref{Tablepropagators}).

The two influence actions we have obtained for the ODF correspond to the effect of the ITB (Eq.(\ref{InfAcQBM})) and the EM field (\eqnref{FEMMA}). Since these environments are independent of each other, their infuence functionals appear as a simple product in the trace over the ODF, i.e.,
\begin{widetext}
\begin{equation}
\begin{split}
\mathcal{F}_{\theta}[\PhDF^{+},\PhDF^{-}]=\int d\PoDFs{\rm f}\int d\PoDFs{\rm in}d\PoDFs{\rm in}'&\rho_{\FPo}(\PoDFs{\rm in},\PoDFs{\rm in}';t_{\rm in})\int^{\PoDFs{\rm f}}_{\PoDFs{\rm in}}\mathcal{D}\PoDF\int^{\PoDFs{\rm f}}_{\PoDFs{\rm in}'}\mathcal{D}\PoDF
\\
&
\times e^{i
\left(S_{\FPo}\left[\PoDF\right]+S_{\rm Int}\left[\PoDF,\PhDF\right]-S_{\FPo}\left[\PoDF'\right]-S_{\rm Int}\left[\PoDF',\PhDF'\right]\right)}\mathcal{F}_{\rm QBM}\left[\PoDF,\PoDF'\right]\mathcal{F}_{\rm EM}^{\rm MA}\left[\PoDF,\PoDF'\right],
\end{split}
\label{InfluenceFunctionalZ}
\end{equation}
%
where $\mathcal{F}_{\rm QBM}\left[\PoDF,\PoDF'\right]\equiv e^{iS_{\rm QBM}[\PoDF,\PoDF']}$. The above expression represents the influence functional modifying the free dynamics of the IDF, due to the dynamics of the ODF which, in turn, evolves under the influence of both the EM field and the ITB.

It is clear that the expression for the influence functional \eqnref{InfluenceFunctionalZ} is given in terms of a Gaussian CTP-integral. It is worth noting that, in this integral, the IDF's variables appear as sources, resulting in $\mathcal{F}_{\theta}$ being a functional of these variables. Moreover, given the form of the actions involved, each cartesian component of the ODF couples only to the same component of the IDF and, as a consequence, the functional integration can be carried out for each component independently. Assuming an initial thermal state for the ODF, the influence functional \eqnref{InfluenceFunctionalZ} can be computed, yielding
\begin{equation}
\mathcal{F}_{\theta}[\PhDF^{+},\PhDF^{-}]=\exp\left\{i
\int_{t_{\rm in}}^{t_{\rm f}} d\lambda''d\lambda'''\PhDF^{-}(\lambda'')\cdot\left[-2\MPh\FPo\FPh\CPoPh^{2}~\PropPo(\lambda''-\lambda''')
~\PhDF^{+}(\lambda''')+\frac{i}{2}\KerNoise_{\rm Env}(\lambda'',\lambda''')~\PhDF^{-}(\lambda''')\right]\right\}.
\label{InfluenceFunctionalPoEM}
\end{equation}
\end{widetext}
Here, 
$\PropPo
$ is the retarded propagator for the ODF under the effective influence of the EM field and the ITB, and obtained from the effective equation of motion of the ODF. In Laplace space, this propagator reads
\begin{equation}\label{eqPropPo}
\PropPo
(s)=\frac{1}{\left[s^{2}+\FPo^{2}+\left(\frac{\CAPo^{2}\FPo^{2}}{6\pi\MPo c^{3}\epsilon_{0}}+4\CDamp\right)s\right]}.
\end{equation}
It is important to remark that it is at this precise point where the Markov approximation discussed above is crucial. Indeed, in order to calculate $\PropPo(s)$ we have employed the Markovian equation of motion Eq.(\ref{MarkovApproxEq}). Contrary to the full equation, the Markovian one, being of second order, has solutions which are fully determined by $\PropPo
(t_{\rm in})$ and $\dot{G}_\Omega
(t_{\rm in})$, as required for dynamical problems. Moreover, under such weak coupling approximation the poles of \eqnref{eqPropPo} have negative real parts, ensuring the causality property and excluding runaways or divergent behaviors.

While in \eqnref{InfluenceFunctionalPoEM} the retarded propagator describes the effective dynamics of the ODF under the action of the EM+ITB environments, the kernel $\KerNoise_{\rm Env}$ contains the information about the fluctuations of ODF, ITB, and EM field as an entire environment, being written as the sum of three contributions,
\begin{equation}
\begin{split}
\KerNoise_{\rm Env}(\lambda'',\lambda''')&=\KerNoise_{\FPo}(\lambda'',\lambda''')+
\\
&+\KerNoise_{\rm EM}(\lambda'',\lambda''')+\KerNoise_{\CDamp}(\lambda'',\lambda'''),
\end{split}
\label{EnvSumKernel}
\end{equation}
each of which is given by
\begin{widetext}
\begin{equation}
    \KerNoise_{\FPo}(\lambda'',\lambda''')=2\MPh\FPh\CPoPh^{2}\coth\left[\frac{\beta_{\FPo}\FPo}{2}\right]\Big[\dot{G}_\Omega
(\lambda''-t_{\rm in})
\dot{G}_\Omega
(\lambda'''-t_{\rm in})+\FPo^{2}\PropPo
(\lambda''-t_{\rm in})\PropPo
(\lambda'''-t_{\rm in})\Big],
\end{equation}
\begin{equation}
    \KerNoise_{\rm EM}(\lambda'',\lambda''')=\frac{2\MPh\FPo\FPh\CPoPh^{2}}{\MPo}\int_{t_{\rm in}}^{t_{\rm f}}d\lambda d\lambda'~\PropPo
(\lambda''-\lambda)
\KerEMH^{\rm MA}(\lambda-\lambda')~\PropPo
(\lambda'''-\lambda'),
\end{equation}
\begin{equation}
    \KerNoise_{\CDamp}(\lambda'',\lambda''')=\frac{2\MPh\FPo\FPh\CPoPh^{2}}{\MPo}\int_{t_{\rm in}}^{t_{\rm f}}d\lambda d\lambda'~\PropPo^{\rm Ret}(\lambda''-\lambda)\NoiQBM(\lambda-\lambda')~\PropPo^{\rm Ret}(\lambda'''-\lambda'),
\end{equation}
\end{widetext}
%
%
where $T_{\FPo}$ is the temperature of the ODF. It is clear that $\KerNoise_{\FPo}$ is related to the initial state of the ODF and its effective dynamics under the influence of the EM+ITB baths, while $\KerNoise_{\rm EM}$ and $\KerNoise_{\CDamp}$ are the noises received by the IDF (through the ODF) from the EM and the ITB environments, respectively.

Finally, note that, from \eqnref{InfluenceFunctionalPoEM}, we can define the influence action for the IDF as
 $\mathcal{F}_{\theta}[\PhDF^{+},\PhDF^{-}]\equiv \exp\{iS_{\rm I\theta}[\PhDF^{+},\PhDF^{-}]\}$, i.e., 
 \begin{equation}
     \begin{split}
         S_{\rm I\theta}[\PhDF^{+},\PhDF^{-}]&=\int_{t_{\rm in}}^{t_{\rm f}}d\lambda d\lambda'\PhDF^{-}(\lambda)\cdot
         \\
         &
         \times
         \Big[-2\MPh\FPo\FPh\CPoPh^{2}\PropPo(\lambda-\lambda')
         \PhDF^{+}(\lambda')+
         \\
         &\hspace{0.6cm}
         +\frac{i}{2}\KerNoise_{\rm Env}(\lambda,\lambda')\PhDF^{-}(\lambda')\Big].
     \end{split}
 \end{equation}
%
This effective action, together with the above defined kernels, allows us to write an effective equation of motion for the IDF as an open system.  Indeed, by extreming the total SCP action of the ODF, $S_{\theta}[\PhDF] - S_{\theta}[\PhDF']+S_{\rm I\theta}[\PhDF^{+},\PhDF^{-}]$, we obtain
\begin{equation}
\ddot{\mathbf{x}}_\omega+\FPh^{2}~\PhDF-2\FPo\FPh\CPoPh^{2}\int_{t_{\rm in}}^{t}d\lambda~\PropPo
(t-\lambda')~\PhDF(\lambda')=0,
\label{EffEqTDF}
\end{equation}
an equation we will use in the following sections in order to calculate the internal energy of the IDF.

\section{The generating functional and the correlation function}\label{GFCF}

This Appendix is devoted to the final step in the calculation of the internal energy of the IDF. First, we will detail the calculation of the generating functional for the IDF. Then, we will use it to calculate the correlations required for the determination of the internal energy. The trace required in this step is schematically depicted in the last column of Table \ref{Tablepropagators}.

The initial definition of the CTP generating functional for the IDF is the extension of \eqnref{FuncionalGeneratriz} to our system:
\begin{widetext}
\begin{equation}\label{GeneratingFunctionalPathIntegral}
    \begin{split}
        Z[\SCTP,\SCTP&']=\int d\PhDFs{\rm f}\int d\PhDFs{\rm in}d\PhDFs{\rm in}'\int d\PoDFs{\rm f}\int d\PoDFs{\rm in}d\PoDFs{\rm in}'\int dA^{\mu}_{\rm f}\int dA^{\mu}_{\rm in}dA'^{\mu}_{\rm in}\prod_{n}\int d\PhBaths{\rm f}\int d\PhBaths{\rm in}d\PhBaths{\rm in}'
        \\
        &
        \times\!\int^{\PhDFs{\rm f}}_{\PhDFs{\rm in}}\mathcal{D}\PhDF\int^{\PhDFs{\rm f}}_{\PhDFs{\rm in}'}\mathcal{D}\PhDF'\int^{\PoDFs{\rm f}}_{\PoDFs{\rm in}}\mathcal{D}\PoDF\int^{\PoDFs{\rm f}}_{\PoDFs{\rm in}'}\mathcal{D}\PoDF'\int^{A^{\mu}_{\rm f}}_{A^{\mu}_{\rm in}}\mathcal{D}A^{\mu}\int^{A^{\mu}_{\rm f}}_{A'^{\mu}_{\rm in}}\mathcal{D}A'^{\mu}\int^{\PhBaths{\rm f}}_{\PhBaths{\rm in}}\mathcal{D}\PhBath\int^{\PhBaths{\rm f}}_{\PhBaths{\rm in}'}\mathcal{D}\PhBath'
        \\
        &
        \times\!\rho\!\left(\PhDFs{\rm in},\PhDFs{\rm in}',\PoDFs{\rm in},\PoDFs{\rm in}',A^{\mu}_{\rm in},A'^{\mu}_{\rm in},\{\PhBaths{\rm in}\},\{\PhBaths{\rm in}'\};t_{\rm in}\right)\!e^{i
        \left(S\left[A^{\mu},\PoDF,\PhDF,\{\PhBath\}\right]-S\left[A'^{\mu},\PoDF',\PhDF',\{\PhBath'\}\right]+\SCTP\ast\PhDF-\SCTP'\ast\PhDF'\right)},
    \end{split}
\end{equation}
where the operation $(\ast)$ is now extended to vector functions as $\mathbf{A}\ast\mathbf{B}\equiv\int^{t_{\rm f}}_{t_{\rm in}}d\tau A_j(\tau)B_j(\tau)$, and the elements of the initial density matrix are defined as 
\begin{equation}
\begin{split}
    \rho(\PhDFs{\rm in},\PhDFs{\rm in}',\PoDFs{\rm in},\PoDFs{\rm in}'&,A^{\mu}_{\rm in},A'^{\mu}_{\rm in},\{\PhBaths{\rm in}\},\{\PhBaths{\rm in}'\};t_{\rm in})\equiv
    \\
    &
    \equiv\langle\PhDFs{\rm in},\PoDFs{\rm in},A^{\mu}_{\rm in},\{\PhBaths{\rm in}\}|\hat{\rho}(t_{\rm in})|\PhDFs{\rm in}',\PoDFs{\rm in}',A'^{\mu}_{\rm in},\{\PhBaths{\rm in}'\}\rangle
    =
    \\
    &
    =\rho_{\rm EM}(A^{\mu}_{\rm in},A'^{\mu}_{\rm in};t_{\rm in})\rho_{\FPo}(\PoDFs{\rm in},\PoDFs{\rm in}';t_{\rm in})\rho_{\theta}(\PhDFs{\rm in},\PhDFs{\rm in}';t_{\rm in})
    \rho_{I}(\{\PhBaths{\rm in}\},\{\PhBaths{\rm in}'\};t_{\rm in})
    .
\end{split}
\end{equation}
\end{widetext}
In the last step above, we have particularized to the case of interest, i.e., that of an uncorrelated initial state.

The functional integrations in \eqnref{GeneratingFunctionalPathIntegral} can be, in principle, performed in any order.
Since we are interested in the dynamics of the IDF, in the previous appendices we have carried out the integrations along the remaining degrees of freedom, i.e.  the ITB (Appendix \ref{TOIDF}), the EM field (Appendix \ref{TEMF}), and the ODF (Appendix \ref{TOODF}). After taking these traces, the generating functional is reduced to
\begin{widetext}
\begin{equation}
Z[\SCTP,\SCTP']=\int d\PhDFs{\rm f}\int d\PhDFs{\rm in}d\PhDFs{\rm in}'~\rho_{\theta}(\PhDFs{\rm in},\PhDFs{\rm in}';t_{\rm in})\int^{\PhDFs{\rm f}}_{\PhDFs{\rm in}}\mathcal{D}\PhDF\int^{\PhDFs{\rm f}}_{\PhDFs{\rm in}'}\mathcal{D}\PhDF'~e^{i
\left(S_{\theta}\left[\PhDF\right]-S_{\theta}\left[\PhDF'\right]+\SCTP\cdot\PhDF-\SCTP'\cdot\PhDF'\right)}\mathcal{F}_{\theta}[\PhDF^{+},\PhDF^{-}],
\label{ZCTPFPH}
\end{equation}
\end{widetext}
where it is evident that the unitary free evolution of the IDF is modified by the environment through its influence functional $\mathcal{F}_{\theta}$. Such influence functional, given in \eqnref{InfluenceFunctionalPoEM}, makes the evolution of the IDF non-unitary, an expected behavior for an open quantum system.

We now proceed to trace out the IDF. Fortunately, the influence functional still conserves a Gaussian shape, with each cartesian component of $\mathbf{x}_\theta$ uncoupled the others. Thus, the integral has the same form as the CTP-integrals performed in the previous sections, and can be carried out analytically. Assuming a thermal state for the IDF, it is straightforward to show that
\begin{equation}
\begin{split}
Z[\SCTP,\SCTP']&= \exp\bigg\lbrace i\int_{t_{\rm in}}^{t_{\rm f}}dtdt'\SCTP^{-}(t)\cdot
\\
&
\times\! \!
\left[\!\PropPh
(t-t')\SCTP^{+}(t')+\frac{i}{2}\CorrPh_{\rm Tot}(t,t')\SCTP^{-}(t')\!\right] \!\!\bigg\rbrace.
\end{split}
\label{GeneratingFunctionalPhDF}
\end{equation}
Here, $\PropPh
$ is the retarded Green function for the IDF under the influence of the composite environment,
and is defined through its Laplace transform as
\begin{equation}
\PropPh
(s)=\frac{1}{\left(s^{2}+\FPh^{2}-2\FPo\FPh\CPoPh^{2}
\PropPo
(s)
\right)}.
\label{PhDFGreen}
\end{equation}
On the other hand, the noise kernel $\CorrPh$ has the same structure as that of the ODF ($\KerNoise_{\rm Env}$), i.e., it is split into different contributions:
\begin{equation}
\CorrPh_{\rm Tot}(t,t')=\CorrPh_{\theta}(t,t')+\CorrPh_{\rm Env}(t,t').
\label{CorrZ}
\end{equation}
The first contribution is related to the initial state of the IDF,
\begin{equation}
\begin{split}
    \CorrPh_{\theta}(t,t')&=\frac{\coth(\beta_\theta\omega_\theta/2)}{2\FPh\MPh}\Big[\dot{\PropPh}(t-t_{\rm in})\dot{\PropPh}
    (t'-t_{\rm in})
    +
    \\
    &+\FPh^{2}\PropPh
    (t-t_{\rm in})\PropPh
    (t'-t_{\rm in})\Big],
    \end{split}
\end{equation}
with $T_{\theta}$ the temperature of the IDF. The second contribution is related to the environment surrounding the IDF, i.e., the rest of the system:
\begin{equation}
    \begin{split}
        \CorrPh_{\rm Env}(t,t')&=\frac{1}{2\MPh^{2}}\int_{t_{\rm in}}^{t_{\rm f}}d\lambda d\lambda'
        \\
        &\times\PropPh
        (t-\lambda)\KerNoise_{\rm Env}(\lambda,\lambda')\PropPh
        (t'-\lambda').
    \end{split}
\end{equation}
The internal structure of $\CorrPh_{\rm Env}$ is inherited from the number of noise kernels perceived by the IDF (see \eqnref{EnvSumKernel}). Note that, whereas the propagator $\PropPh$ has a clear interpretation as a retarded kernel, the kernel $\CorrPh_{\rm Env}$ here represents the contribution to the correlations of the IDF originated in the fluctuations of the environment.


Once we have calculated the generating functional $Z$, we can obtain any quantum correlation of the IDF by taking functional derivatives (see \eqnref{G12derivatives}):
\begin{equation}
\left\langle\hat{x}_\theta^{j}(t)\hat{x}_\theta^{k}(t')\right\rangle=\frac{\delta^{2}Z}{\delta\SCTPc^{j}(t)\delta\SCTPc'^{k}(t')}.
\label{CorrZRelation}
\end{equation}
We are interested in the expected value of the bare energy of the IDF, i.e., $\hat{H}_{\theta}=(\MoPh^{2}/2\MPh)+(\MPh\FPh^{2}\hat{\mathbf{x}}_\theta^{2}/2)$. Thus, we need to calculate two correlators, namely $\langle \hat{\mathbf{x}}_\theta^2\rangle$ and $\langle \hat{\mathbf{p}}_\theta^2\rangle$. The former is straightforward to compute using the above equation, as 
\begin{equation}
\begin{split}
    \langle\hat{\mathbf{x}}_\theta^2(t)\rangle &= \lim_{t'\rightarrow t}\delta_{jk}\left\langle\hat{x}_\theta^{j}(t)\hat{x}_\theta^{k}(t')\right\rangle
    \\
    &
    =
    \lim_{t'\rightarrow t}\frac{\delta_{jk}}{2}\left\langle\hat{x}_\theta^{j}(t)\hat{x}_\theta^{k}(t')+\hat{x}_\theta^{k}(t')\hat{x}_\theta^{j}(t)\right\rangle.
\end{split}
\end{equation}
Using the final form of the generating functional Eq. (\ref{GeneratingFunctionalPhDF}), the above quantum correlation of the IDF can be calculated in a straightforward way  \cite{CalzettaRouraVerdaguer}, yielding
\begin{equation}
\frac{1}{2}\langle\hat{\PhDFc}^{j}(t)\hat{\PhDFc}^{k}(t')+\hat{\PhDFc}^{k}(t')\hat{\PhDFc}^{j}(t)\rangle=\delta_{jk}\CorrPh_{\rm Tot}(t,t').
\end{equation}
By substitution of the noise kernel \eqnref{CorrZ} in the above expression, we find
\begin{widetext}
\begin{equation}\label{expectedxx}
    \begin{split}
        \langle\hat{\mathbf{x}}_\theta^{2}&(t)\rangle=3\CorrPh_{\rm Tot}(t,t)=3\CorrPh_{\theta}(t,t)+3\CorrPh_{\rm Env}(t,t)=
        \\
        &
        =\frac{3}{2\MPh^{2}}\!\left[\frac{\MPh}{\FPh}\!\coth\!\!\left[\frac{\beta_{\theta}\FPh}{2}\right]\!\!\left(\!\left[\dot{\PropPh}
        (t-t_{\rm in})\right]^{2}+\FPh^{2}\left[\PropPh
        (t-t_{\rm in})\right]^{2}\right)\!+\!\int_{t_{\rm in}}^{t_{\rm f}}\!\!d\lambda d\lambda'\PropPh
        (t-\lambda)\KerNoise_{\rm Env}(\lambda,\lambda')\PropPh
        (t-\lambda')\right]\!.
    \end{split}
\end{equation}
\end{widetext}
We now focus on the correlator between momenta. Note that, since all the coupling terms related with the IDF contain only the degree of freedom $\mathbf{x}_\theta$ and not its time derivatives, the canonical momentum is defined as usual, $\MoPh=\MPh\dot{\hat{\mathbf{x}}}_\theta$. Thus, the expected value we seek is proportional to the correlation $\langle\dot{\hat{\mathbf{x}}}_\theta^2(t)\rangle$, which we can calculate as
\begin{widetext}
\begin{equation}\label{expectedxdotxdot}
\begin{split}
    \left\langle\right.\dot{\hat{\mathbf{x}}}_\theta^{2}&(t)\left.\!\right\rangle=\lim_{t'\rightarrow t}\delta_{jk}\left\langle\dot{\hat{x}}_\theta^{j}(t)\dot{\hat{x}}_\theta^{k}(t')\right\rangle 
    =\lim_{t'\rightarrow t}\frac{\delta_{jk}}{2}\partial_{t}\partial_{t'}\langle\hat{x}_\theta^{j}(t)\hat{x}_\theta^{k}(t')+\hat{x}_\theta^{k}(t')\hat{x}_\theta^{j}(t)\rangle=\lim_{t'\rightarrow t}\frac{3}{2}\partial_{t}\partial_{t'}\CorrPh_{\rm Tot}(t,t')
    \\
    &
    =
    \frac{3}{2\FPh\MPh}\coth\!\!\left[\frac{\beta_{\theta}\FPh}{2}\right]\!\!\left[\!\left(\ddot{\PropPh}
    (t-t_{\rm in})\right)^{2}\!+\FPh^{2}\left(\dot{\PropPh}
    (t-t_{\rm in})\right)^{2}\right]\!+\!\frac{3}{2\MPh^{2}}\!\int_{t_{\rm in}}^{t_{\rm f}}\!\!d\lambda d\lambda'\dot{\PropPh}
    (t-\lambda)\KerNoise_{\rm Env}(\lambda,\lambda')\dot{\PropPh}
    (t-\lambda')
    .
\end{split}
\end{equation}
In the last line, we have taken into account the causal property of the retarded Green function, $\PropPh \propto \theta(t-t')$, that restricts the integration interval, as well as the fact that $t,t'>t_{\rm in}$.

Finally, combining \eqnref{expectedxx} and \ref{expectedxdotxdot}, we can immediately determine the expected value of the internal energy as
\begin{equation}
\begin{split}
    \langle\hat{H}_{\theta}(t)\rangle&=\frac{\MPh}{2}\left\langle\dot{\hat{\mathbf{x}}}_\theta^{2}(t)\right\rangle+\frac{\MPh\FPh^{2}}{2}\left\langle\hat{x}_\theta^{2}(t)\right\rangle=\lim_{t'\rightarrow t}\frac{\MPh}{2}\left[\partial_{t}\partial_{t'}+\FPh^{2}\right]\frac{\delta_{jk}}{2}\left\langle\hat{x}_\theta^{j}(t)\hat{x}_\theta^{k}(t')+\hat{x}_\theta^{k}(t')\hat{x}_\theta^{j}(t)\right\rangle
    \\
    &=\frac{3}{4\FPh}\coth\left(\frac{\beta_{\theta}\FPh}{2}\right)\left(\left[\ddot{\PropPh}
    (t-t_{\rm in})\right]^{2}+2\FPh^{2}\left[\dot{\PropPh}
    (t-t_{\rm in})\right]^{2}+\FPh^{4}\left[\PropPh(t-t_{\rm in})\right]^{2}\right)
    \\
    &+\frac{3}{4\MPh}\int_{t_{\rm in}}^{t}d\lambda d\lambda'\Big[\dot{\PropPh}(t-\lambda)\KerNoise_{\rm Env}(\lambda,\lambda')\dot{\PropPh}
    (t-\lambda')+\FPh^{2}~\PropPh
    (t-\lambda)\KerNoise_{\rm Env}(\lambda,\lambda')\PropPh
    (t-\lambda')\Big],
\end{split}\label{HTDFFull}
\end{equation}
\end{widetext}
which corresponds to Eq.(\ref{EnergyNormalizedTime}) after setting $t_{\rm in}=0$. This expression describes the full time evolution of the energy of the IDF. Note that due to the properties of the retarded propagators, the initial energy is $\langle\hat{H}_{\theta}(t_{\rm in})\rangle=(3/2)\FPh\coth\left(\beta_{\theta}\FPh/2\right)$. As expected from our choice of initial state, this is the energy of a harmonic oscillator in a thermal state.

\section{Limiting behaviors of the IDF's energy}\label{LBTDFE}

This appendix is devoted to deduce the short- and long-time asymptotic expressions of the internal energy Eq.(\ref{HTDFFull}) or, equivalently, of  Eq.(\ref{EnergyNormalizedTime}). Specifically, we aim at recovering  Eqs.(\ref{shorttime}) and (\ref{Hinfinity}).

\subsection{Short-time behavior}

Let us start by noticing that, from their Laplace transforms, both the time-domain Green function of the IDF and its time derivatives can be calculated by residues through a Mellin formula,
\begin{equation}
\PropPh(t)=\!\!\int_{l-i\infty}^{l+i\infty}\!\!\frac{ds}{2\pi i}e^{st}\PropPh(s)=\!\!
\int_{l-i\infty}^{l+i\infty}\!\!\frac{dv}{2\pi i}e^{v}\PropPh\!\left(\frac{v}{t}\right)\!.
\label{MellinForExpansion}
\end{equation}
The convenient change of variable performed in the last step above will allow us to expand the internal energy for short times, $t\gtrapprox 0$.
We start by expanding the first contribution in Eq.(\ref{HTDFFull}), i.e., the terms associated to the initial state of the IDF:
\begin{widetext}
\begin{equation}\label{shorttimes1}
\frac{3}{4\FPh}\coth\left[\frac{\beta_{\theta}\FPh}{2}\right]\Big[\left[\ddot{\PropPh}
(t-t_{\rm in})\right]^{2}+2\FPh^{2}\left[\dot{\PropPh}
(t-t_{\rm in})\right]^{2}+\FPh^{4}\left[\PropPh
(t-t_{\rm in})\right]^{2}\Big]\approx\frac{3\FPh}{2}\coth\left(\frac{\beta_{\theta}\FPh}{2}\right)
+\mathcal{O}\left[(t-t_{\rm in})^{4}\right].
\end{equation}
\end{widetext}
Note that the above term is zero up to fourth order in time, a direct consequence of the initial thermal state commuting with the free Hamiltonian of the IDF. The time evolution at short times will thus be generated only by the interaction terms, given by the second line of \eqnref{HTDFFull}. In order to calculate such second term, we write the integrands as a two-times-coincidence limit ($t'\rightarrow t$) of a correlation function depending on $t,t'$, i.e., $\PropPh
    (t-\lambda)\KerNoise_{\rm Env}(\lambda,\lambda')\PropPh
    (t-\lambda') = \lim_{t\to t'}\PropPh
    (t-\lambda)\KerNoise_{\rm Env}(\lambda,\lambda')\PropPh
    (t'-\lambda')$.
Then, we double-Laplace transform in each variable $t,t'$, and perform the change of variables shown in Eq.(\ref{MellinForExpansion}) in both Laplace integrals. We then expand each integral for short times and, as a final step, take the limit $t\to t'$ to obtain
\begin{widetext}
\begin{equation}\label{shorttimes2}
\begin{split}
    \frac{3}{4\MPh}\int_{t_{\rm in}}^{t}d\lambda d\lambda'\Big[\dot{\PropPh}
    (t-\lambda)\KerNoise_{\rm Env}(\lambda,\lambda')&\dot{\PropPh}
    (t-\lambda')+\FPh^{2}~\PropPh
    (t-\lambda)\KerNoise_{\rm Env}(\lambda,\lambda')\PropPh
    (t-\lambda')\Big]\approx
    \\
    &
    \approx\frac{3}{2}\FPh\CPoPh^{2}\coth\left[\frac{\beta_{\FPo}}{2}\FPo\right](t-t_{\rm in})^{2}\left[1-\left(4\CDamp+\frac{\CAPo^{2}\FPo^{2}}{6\pi\MPo c^{3}\epsilon_{0}}\right)\frac{(t-t_{\rm in})}{2}\right],
\end{split}
\end{equation}
up to third order. 
%
%
%
%
%
Finally, by combining \eqnref{shorttimes1} and \ref{shorttimes2}, we can express the expectation value of the energy at short times as
\begin{equation}\label{Hshorttime}
    \left\langle\hat{H}_{\theta}(t)\right\rangle\approx\frac{3\FPh}{2}\coth\left[\frac{\beta_{\theta}\FPh}{2}\right]+\frac{3}{2}\FPh\CPoPh^{2}\coth\left[\frac{\beta_{\FPo}}{2}\FPo\right](t-t_{\rm in})^{2}\left[1-\left(4\CDamp+\frac{\CAPo^{2}\FPo^{2}}{6\pi\MPo c^{3}\epsilon_{0}}\right)\frac{(t-t_{\rm in})}{2}\right].
\end{equation}
\end{widetext}
Note that, in the equation above, at short times $\mathcal{O}\left((t-t_{\rm in})^2\right)$ only the information of the ODF appears. On the other hand, the fluctuations of both the EM field and the ITB only act on the IDF at a later step $\mathcal{O}\left((t-t_{\rm in})^3\right)$. This third order correction is always negative, thus reverting the universal quadratic growth to a physically consistent decay of the internal energy. As a final remark, note that by setting $t_{\rm in}=0$, \eqnref{Hshorttime} can be easily cast in the form of Eq.(\ref{shorttime}) in the main text.

\subsection{Long-time limit}

In order to obtain the long time limit of Eq. (\ref{HTDFFull}), we take the limit $t_{\rm in}\rightarrow-\infty$.
For each of the terms appearing in Eq.(\ref{HTDFFull}), such limit is calculated in different ways. We start by the first line, related to the initial state of the IDF. Here, we use the \emph{final value theorem}, which states that the long-time limit of a given time-dependent function $f(t)$ is equal to $\lim_{s\to 0}sf(s)$, $f(s)$ being the Laplace transform of $f(t)$. Noting that the limit of each term as $s\to 0$ is finite, we immediately find that
\begin{equation}
\begin{split}
    \lim_{t_{\rm in}\rightarrow-\infty}\Big(\left[\ddot{\PropPh}
    (t-t_{\rm in})\right]^{2}&+2\FPh^{2}\left[\dot{\PropPh}
    (t-t_{\rm in})\right]^{2}+
    \\
    &
    +\FPh^{4}\left[\PropPh
    (t-t_{\rm in})\right]^{2}\Big)=0.
\end{split}
\end{equation}
In other words, the asymptotic state of the system for long times has no memory of the initial state of the IDF.

Regarding the second term in Eq. (\ref{HTDFFull}), it is more convenient to take its long time limit directly in the integral. Note that, by the same argument given above, it is straightforward to show that $\KerNoise_{\FPo}\rightarrow 0$ for $t_{\rm in}\rightarrow-\infty$. Thus, we have
\begin{widetext}
\begin{equation}
\begin{split}
    &\int_{t_{\rm in}}^{t}d\lambda     d\lambda'\Big[\dot{\PropPh}
    (t-\lambda)\KerNoise_{\rm         Env}(\lambda,\lambda')\dot{\PropPh}
    (t-\lambda')+\FPh^{2}~\PropPh
    (t-\lambda)\KerNoise_{\rm Env}(\lambda,\lambda')\PropPh
    (t-\lambda')\Big]\rightarrow
    \\
    &
    \rightarrow\int_{-\infty}^{+\infty}\!\!\!\!d\lambda d\lambda'\Big[\dot{\PropPh}
    (t-\lambda)\!\left[\KerNoise_{\rm EM}(\lambda,\lambda')\!+\!\KerNoise_{\CDamp}(\lambda,\lambda')\right]\!\dot{\PropPh}
    (t-\lambda')\!+\!\FPh^{2}\PropPh
    (t-\lambda)\!\left[\KerNoise_{\rm EM}(\lambda,\lambda')\!+\!\KerNoise_{\CDamp}(\lambda,\lambda')\right]\!\PropPh
    (t-\lambda')\Big],
\end{split}
\end{equation}
where we have used the causal property of the Green functions to extend the upper limit of the integral to $+\infty$.
Moreover, such causal behavior allows us to compute the above expression in Fourier space since, for any causal time-dependent function $f(t)$, the Fourier and Laplace transforms are related by $\overline{f}(\omega)=f(s=-i\omega)$. Then, using the expressions of the Fourier transforms $\FouKerEMH^{\rm MA}$ (Eq.(\ref{FDRMA})) and $\overline{\NoiQBMNS}_{\gamma}$  (Eq.(\ref{NoiseQBMDef})), we can calculate the value on the r.h.s. of the last equation as a single integral in frequency space,
\begin{equation}
\begin{split}
    \int_{t_{\rm in}}^{t}&d\lambda d\lambda'\Big[\dot{\PropPh}
    (t-\lambda)\KerNoise_{\rm Env}(\lambda,\lambda')\dot{\PropPh}
    (t-\lambda')+\FPh^{2}~\PropPh
    (t-\lambda)\KerNoise_{\rm Env}(\lambda,\lambda')\PropPh
    (t-\lambda')\Big]\longrightarrow
    \\
    &
    \rightarrow 4\MPh\FPh\CPoPh^{2}\FPo\int_{0}^{+\infty}\frac{d\omega}{2\pi}\omega\left[\frac{\CAPo^{2}\FPo^{2}}{6\pi\MPo c^{3}\epsilon_{0}}\coth\left(\frac{\beta_{\rm EM}}{2}\omega\right)+4\CDamp\coth\left(\frac{\beta_{\CDamp}}{2}\omega\right)\right](\omega^{2}+\FPh^{2})\left|
    \PropPh
    (-i\omega)
    \PropPo
    (-i\omega)\right|^{2}.
\end{split}
\end{equation}
Here, we have 
omitted the cut-off functions $\EMfCO(\omega),\FfCO(\omega)$ for simplicity, although we will introduce them when computing the integral. From the equation above, it is straightforward to write the asymptotic long-time limit of the internal energy as
\begin{equation}
\langle\hat{H}_{\theta}^{\infty}\rangle=3\FPh\CPoPh^{2}\FPo\int_{0}^{+\infty}\frac{d\omega}{2\pi}~\omega\left(\frac{\CAPo^{2}\FPo^{2}}{6\pi\MPo c^{3}\epsilon_{0}}\coth\left[\frac{\beta_{\rm EM}}{2}\omega\right]+4\CDamp\coth\left[\frac{\beta_{\CDamp}}{2}\omega\right]\right)(\omega^{2}+\FPh^{2})\left|
\PropPh
(-i\omega)\right|^{2}\left|
\PropPo
(-i\omega)\right|^{2},
\label{LongTimeLimitHz}
\end{equation}
\end{widetext}
which can easily be recast into the form of \eqnref{Hinfinity}.
In the above equation, it is clear that the only contributions stem from the EM and ITB environments, which provide fluctuations to the IDF. 

%

 
\bibliographystyle{apsrev4-1}
\bibliography{bibliography}

\begin{thebibliography}{35}
\bibitem [{foo()}]{footnote1}%
  \BibitemOpen
  \href@noop {} {}\bibinfo {note} {Note that, although mentioned here, the
  condition $k_{\rm B}T_{\rm EM} /\hbar \ll \Omega$ is not an assumption of our
  model, which remains valid for general materials. For semiconductors, where
  $k_{\rm B}T_{\rm EM} /\hbar \approx \Omega$, our model still applies, but the
  derivations in Sec. IV require some modifications. Specifically, the matching
  with the polarizability could be attained by considering a multiple resonance
  permittivity, and the definition of the internal energy should include a
  contribution from the ODF. In any case, however, the multiexponential
  behavior of Fig. \ref{figTimeEvol}, and thus the discrepancy with FED,
  remains.}\BibitemShut {Stop}%
\bibitem [{not()}]{notecutoffs}%
  \BibitemOpen
  \href@noop {} {}\bibinfo {note} {For all the calculations shown in the
  figures, we have used Lorentzian cutoff functions for both the EM and the ITB
  baths. Regarding the EM field, the cutoff frequency is set to $\Lambda_{\rm
  EM} = m_ec^2/\hbar$, where $m_e$ is the electron mass. On the other hand, for
  the ITB we use $\Lambda_{\rm I} = N\omega_\text{phon}$, where $N$ is the
  total number of atoms in the NP.}\BibitemShut {Stop}%
\end{thebibliography}

\end{document}